\documentclass[zpreprint,zbstnp]{zeus_paper}

\usepackage[english]{babel}

\newcommand{\Zdetdesc}{%
A detailed description of the ZEUS detector can be found 
elsewhere~\cite{zeus:1993:bluebook}. A brief outline of the 
components that are most relevant for this analysis is given
below.\xspace}

\newcommand{\Zcaldesc}{%
The high-resolution uranium--scintillator calorimeter (CAL)~\citeCAL consists 
of three parts: the forward (FCAL), the barrel (BCAL) and the rear (RCAL)
calorimeters. Each part is subdivided transversely into towers and
longitudinally into one electromagnetic section (EMC) and either one (in RCAL)
or two (in BCAL and FCAL) hadronic sections (HAC). The smallest subdivision of
the calorimeter is called a cell.  The CAL energy resolutions, as measured under
test-beam conditions, are $\sigma(E)/E=0.18/\sqrt{E}$ for electrons and
$\sigma(E)/E=0.35/\sqrt{E}$ for hadrons ($E$ in $\Gev$).}

\chardef\usc=95
\chardef\til=126
\catcode`\@=11 
\DeclareRobustCommand\xdotspace{\futurelet\@let@token\@xdotspace}
\def\@xdotspace{%
  \ifx\@let@token.\else
  \ifx\@let@token\bgroup.\else
  \ifx\@let@token\egroup.\else
  \ifx\@let@token\/.\else
  \ifx\@let@token\ .\else
  \ifx\@let@token~.\else
  \ifx\@let@token!.\else
  \ifx\@let@token,.\else
  \ifx\@let@token:.\else
  \ifx\@let@token;.\else
  \ifx\@let@token?.\else
  \ifx\@let@token/.\else
  \ifx\@let@token'.\else
  \ifx\@let@token).\else
  \ifx\@let@token-.\else
  \ifx\@let@token\@xobeysp.\else
  \ifx\@let@token\space.\else
  \ifx\@let@token\@sptoken.\else
   .\space
   \fi\fi\fi\fi\fi\fi\fi\fi\fi\fi\fi\fi\fi\fi\fi\fi\fi\fi}
\catcode`\@=12 

\newcommand{\stru}[2]{%
   \relax\ifmmode\hbox{\vrule height#1 depth#2 width0pt}%
   \else\vrule height#1 depth#2 width0pt\fi}

\newcommand{\Ronum}[1]{\uppercase\expandafter{\romannumeral#1}}
\newcommand{\ronum}[1]{\expandafter{\romannumeral#1}}
\DeclareRobustCommand{\LaTeXZ}{%
  \LaTeX\kern-.05em4\kern-.1em
  {\raisebox{-0.2ex}{$\scriptstyle\text{ZEUS}$}}\xspace}

\newcommand{\fig}[1]{Fig.~\ref{fig-#1}}
\newcommand{\Fig}[1]{Figure~\ref{fig-#1}}
\newcommand{\figand}[2]{Figs.~\ref{fig-#1} and~\ref{fig-#2}}

\newcommand{\tab}[1]{Table~\ref{tab-#1}}

\newcommand{\Sect}[1]{Section~\ref{sec-#1}}

\DeclareMathAlphabet{\mathbf}{OT1}{cmr}{bx}{sl}
\newcommand{\eVdist}{\kern-0.06667em}

\newcommand{\Gev}{{\text{Ge}\eVdist\text{V\/}}}

\newcommand{\mev}{{\,\text{Me}\eVdist\text{V\/}}}
\newcommand{\gev}{{\,\text{Ge}\eVdist\text{V\/}}}

\newcommand{\pb}{\,\text{pb}}

\newcommand{\nbi}{\,\text{nb}^{-1}}
\newcommand{\pbi}{\,\text{pb}^{-1}}

\newcommand{\met}{\,\text{m}}

\newcommand{\cm}{\,\text{cm}}

\newcommand{\mrad}{\,\text{mrad}}

\newcommand{\slashfrac}[2]{%
  \raisebox{0.5ex}{\ensuremath #1}\kern-0.12em/\kern-0.08em
  \raisebox{-.8ex}{\ensuremath #2}}

\newcommand{\sqr}[3]{%
    {\vcenter{\hrule height.#3ex\hbox{\vrule width.#2ex height#1ex
     \kern#1ex\vrule width.#3ex}\hrule height.#2ex}}}

\catcode`\@=11 
\newcommand{\parenbar}{\mathpalette\p@renb@r}
\def\p@renb@r#1#2{\vbox{%
  \ifx#1\scriptscriptstyle \dimen@.7em\dimen@ii.2em\else
  \ifx#1\scriptstyle \dimen@.8em\dimen@ii.25em\else
  \dimen@1em\dimen@ii.4em\fi\fi \offinterlineskip
  \ialign{\hfill##\hfill\cr
    \vbox{\hrule width\dimen@ii}\cr
    \noalign{\vskip-.3ex}%
    \hbox to\dimen@{$\mathchar300\hfil\mathchar301$}\cr
    \noalign{\vskip-.3ex}%
    $#1#2$\cr}}}
\catcode`\@=12 

\newcommand{\IP}{{\rm I$\kern-0.01667em$P}\xspace}

\mathchardef\qsm=63
\mathchardef\pls=43
\mathchardef\mns=512
\mathchardef\plm=518
\mathchardef\eql=61
\mathchardef\smallleft=300
\mathchardef\smallright=301
\mathchardef\les=316
\mathchardef\gre=318
\mathchardef\leq=532
\mathchardef\grq=533

\catcode`\@=11 
\newcounter{pict@width}
\newcounter{pict@height}
\newlength{\pict@scale}
\newcommand{\psfigadd}[4]{%
\setcounter{pict@width}{1*\ratio{#2+\pict@scale/2}{\pict@scale}}
\setcounter{pict@height}{1*\ratio{#3+\pict@scale/2}{\pict@scale}}
\setlength{\unitlength}{\pict@scale}
\hbox to #2{\hspace{-\fill}\begin{picture}(\thepict@width,\thepict@height)
\put(0,0){\psfig{figure=#1,width=#2,height=#3,clip=}}
\SetScale{0.283466457}
\SetWidth{1.763889}
{#4}
\end{picture}}
}
\newcounter{pict@widthfst}
\newcounter{pict@widthscd}
\newcounter{pict@widthtot}
\newcommand{\psfigaddtwo}[7]{%
\setcounter{pict@widthfst}{1*\ratio{#2+\pict@scale/2}{\pict@scale}}
\setcounter{pict@widthscd}{1*\ratio{#2+#4+\pict@scale/2}{\pict@scale}}
\setcounter{pict@widthtot}{1*\ratio{#2+#4+#6+\pict@scale/2}{\pict@scale}}
\setcounter{pict@height}{1*\ratio{#3+\pict@scale/2}{\pict@scale}}
\setlength{\unitlength}{\pict@scale}
\hbox{\hspace{-\fill}\begin{picture}(\thepict@widthtot,\thepict@height)
\put(0,0){\psfig{figure=#1,width=#2,height=#3,clip=}}
\put(\thepict@widthscd,0){\psfig{figure=#5,width=#6,height=#3,clip=}}
\SetScale{0.283466457}
\SetWidth{1.763889}
{#7}
\end{picture}}
}
\newcommand{\psfigror}[4]{%
\setcounter{pict@width}{1*\ratio{#2+\pict@scale/2}{\pict@scale}}
\setcounter{pict@height}{1*\ratio{#3+\pict@scale/2}{\pict@scale}}
\setlength{\unitlength}{\pict@scale}
\hbox{\begin{picture}(\thepict@width,\thepict@height)
\put(0,\thepict@height){\psfig{figure=#1,width=#3,height=#2,clip=,angle=270}}
\SetScale{0.283466457}
\SetWidth{1.763889}
{#4}
\end{picture}}
}
\newcommand{\psfigrol}[4]{%
\setcounter{pict@width}{1*\ratio{#2+\pict@scale/2}{\pict@scale}}
\setcounter{pict@height}{1*\ratio{#3+\pict@scale/2}{\pict@scale}}
\setlength{\unitlength}{\pict@scale}
\hbox{\begin{picture}(\thepict@width,\thepict@height)
\put(0,0){\psfig{figure=#1,width=#3,height=#2,clip=,angle=90}}
\SetScale{0.283466457}
\SetWidth{1.763889}
{#4}
\end{picture}}
}
\catcode`\@=12 
\newlength\listtextwidth

\newcommand{\pcite}[1]{{\protect\cite{#1}}}

\catcode`\@=11 
\newlength{\@tabfninsert}
\newlength{\@tabfnwidth}
\newcommand{\tabfootnote}[2]{%
  \setlength{\@tabfninsert}{0.8em}
  \setlength{\@tabfnwidth}{\textwidth}
  \addtolength{\@tabfnwidth}{-\@tabfninsert}
  \addtolength{\@tabfnwidth}{-0.4em}
  \noindent\makebox[\@tabfninsert][r]{\footnotesize$^{#1}$\hfil}\hfill%
  \parbox[t]{\@tabfnwidth}{\footnotesize #2\hfill}}
\catcode`\@=12 

\newcommand{\myZcoosysA}{%
The ZEUS coordinate system is a right-handed Cartesian system, with the $Z$
axis pointing in the proton beam direction, referred to as the ``forward
direction'', and the $X$ axis pointing
towards the center of HERA.
The coordinate origin is at the nominal interaction point.\xspace}
\newcommand{\myZcoosysfnA}{\footnote{\myZcoosysA}}
\def\NSZ{hep-ph-9708290}
\def\DAP{epj:a7:109}
\def\citeCAL{{\cite{%
nim:a309:77,*nim:a309:101,*nim:a321:356,*nim:a336:23%
}}\xspace}

\begin{document}

\prepnum{{DESY--07--011}}

\title{
Leading neutron energy and $\mathbf{p_T}$\\
distributions in deep inelastic scattering\\
and photoproduction at HERA 
}

\author{ZEUS Collaboration}
\date{February 2007}

\abstract{
The production of energetic neutrons in $ep$ collisions has been studied
with the ZEUS detector at HERA.
The neutron energy and $p_T^2$ distributions
were measured with a forward neutron calorimeter and tracker
in a $40 \pb^{-1}$ sample of
inclusive deep inelastic scattering (DIS) data
and a $6 \pb^{-1}$ sample of photoproduction data.
The neutron yield in photoproduction is suppressed relative
to DIS for the lower neutron energies and the neutrons have a steeper
$p_T^2$ distribution,
consistent with the expectation from absorption models.
The distributions are compared
to HERA
measurements of leading protons.
The neutron energy and transverse-momentum distributions in DIS are compared
to Monte Carlo simulations and
to the predictions of particle exchange models.
Models of pion exchange incorporating absorption and additional
secondary meson exchanges give a good description of the data.
}

\makezeustitle

\def\3{\ss}

%
%
%
%
\pagenumbering{Roman}                                                                               
\begin{center}                                                                                     
{                      \Large  The ZEUS Collaboration              }                               
\end{center}                                                                                       
  S.~Chekanov$^{   1}$,                                                                            
  M.~Derrick,                                                                                      
  S.~Magill,                                                                                       
  S.~Miglioranzi$^{   2}$,                                                                         
  B.~Musgrave,                                                                                     
  D.~Nicholass$^{   2}$,                                                                           
  \mbox{J.~Repond},                                                                                
  R.~Yoshida\\                                                                                     
 {\it Argonne National Laboratory, Argonne, Illinois 60439-4815}, USA~$^{n}$                       
\par \filbreak                                                                                     
  M.C.K.~Mattingly \\                                                                              
 {\it Andrews University, Berrien Springs, Michigan 49104-0380}, USA                               
\par \filbreak                                                                                     
  M.~Jechow, N.~Pavel~$^{\dagger}$, A.G.~Yag\"ues Molina \\                                        
  {\it Institut f\"ur Physik der Humboldt-Universit\"at zu Berlin,                                 
           Berlin, Germany}
% ...
\newpage                                                                                           
$^{\    1}$ supported by DESY, Germany \\                                                          
$^{\    2}$ also affiliated with University College London, UK \\                                  
% ...
$^{\dagger}$ deceased \\                                                                           
%
\newpage   
                                                           %
                                                           %
\begin{tabular}[h]{rp{14cm}}                                                                       
$^{a}$ &  supported by the Natural Sciences and Engineering Research Council of Canada (NSERC) \\  
$^{b}$ &  supported by the German Federal Ministry for Education and Research (BMBF), under        
          contract numbers HZ1GUA 2, HZ1GUB 0, HZ1PDA 5, HZ1VFA 5\\                                
$^{c}$ &  supported in part by the MINERVA Gesellschaft f\"ur Forschung GmbH, the Israel Science   
          Foundation (grant no. 293/02-11.2) and the U.S.-Israel Binational Science Foundation \\  
$^{d}$ &  supported by the German-Israeli Foundation and the Israel Science Foundation\\           
$^{e}$ &  supported by the Italian National Institute for Nuclear Physics (INFN) \\                
$^{f}$ &  supported by the Japanese Ministry of Education, Culture, Sports, Science and Technology 
          (MEXT) and its grants for Scientific Research\\                                          
$^{g}$ &  supported by the Korean Ministry of Education and Korea Science and Engineering          
          Foundation\\                                                                             
$^{h}$ &  supported by the Netherlands Foundation for Research on Matter (FOM)\\                   
$^{i}$ &  supported by the Polish State Committee for Scientific Research, grant no.               
          620/E-77/SPB/DESY/P-03/DZ 117/2003-2005 and grant no. 1P03B07427/2004-2006\\             
$^{j}$ &  partially supported by the German Federal Ministry for Education and Research (BMBF)\\   
$^{k}$ &  supported by RF Presidential grant N 8122.2006.2 for the leading                         
          scientific schools and by the Russian Ministry of Education and Science through its grant
          Research on High Energy Physics\\                                                        
$^{l}$ &  supported by the Spanish Ministry of Education and Science through funds provided by     
          CICYT\\                                                                                  
$^{m}$ &  supported by the Particle Physics and Astronomy Research Council, UK\\                   
$^{n}$ &  supported by the US Department of Energy\\                                               
$^{o}$ &  supported by the US National Science Foundation. Any opinion,                            
findings and conclusions or recommendations expressed in this material                             
are those of the authors and do not necessarily reflect the views of the                           
National Science Foundation.\\                                                                     
$^{p}$ &  supported by the Polish Ministry of Science and Higher Education\\                       
$^{q}$ &  supported by FNRS and its associated funds (IISN and FRIA) and by an Inter-University    
          Attraction Poles Programme subsidised by the Belgian Federal Science Policy Office\\     
$^{r}$ &  supported by the Malaysian Ministry of Science, Technology and                           
Innovation/Akademi Sains Malaysia grant SAGA 66-02-03-0048\\                                       
\end{tabular}                                                                                     
\newpage

\pagenumbering{arabic} 
\pagestyle{plain}

\section{Introduction}
\label{sec-int}

\begin{figure}[b]
\vspace{-1.0cm}
\begin{tabular}{cccc}
 a)
&
 \begin{minipage}{7cm}
\epsfig{file=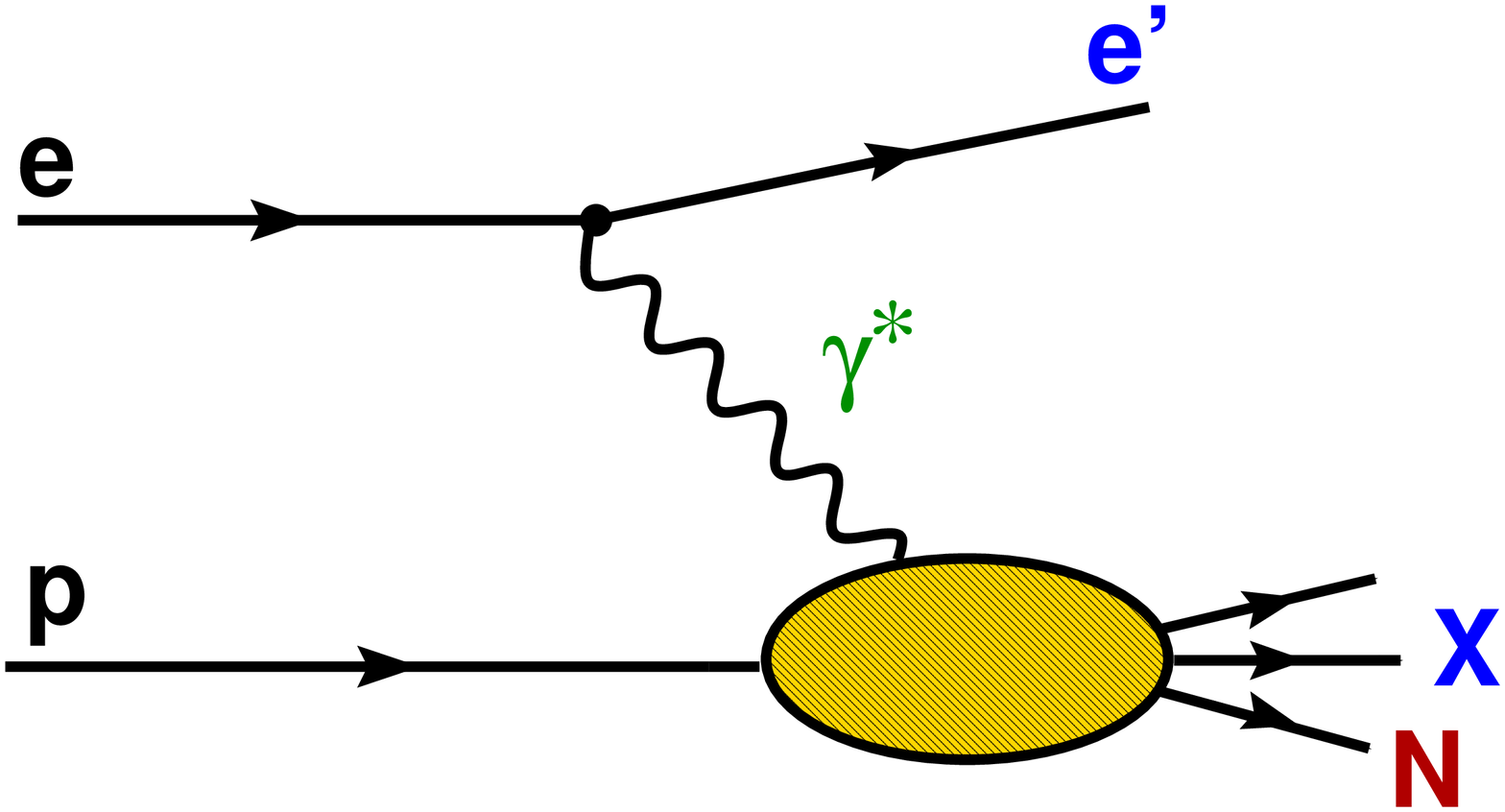,clip=,,width=6cm}
 \end{minipage}
 &
 b)
&
 \begin{minipage}{7cm}
\vspace{0.2cm}
\epsfig{file=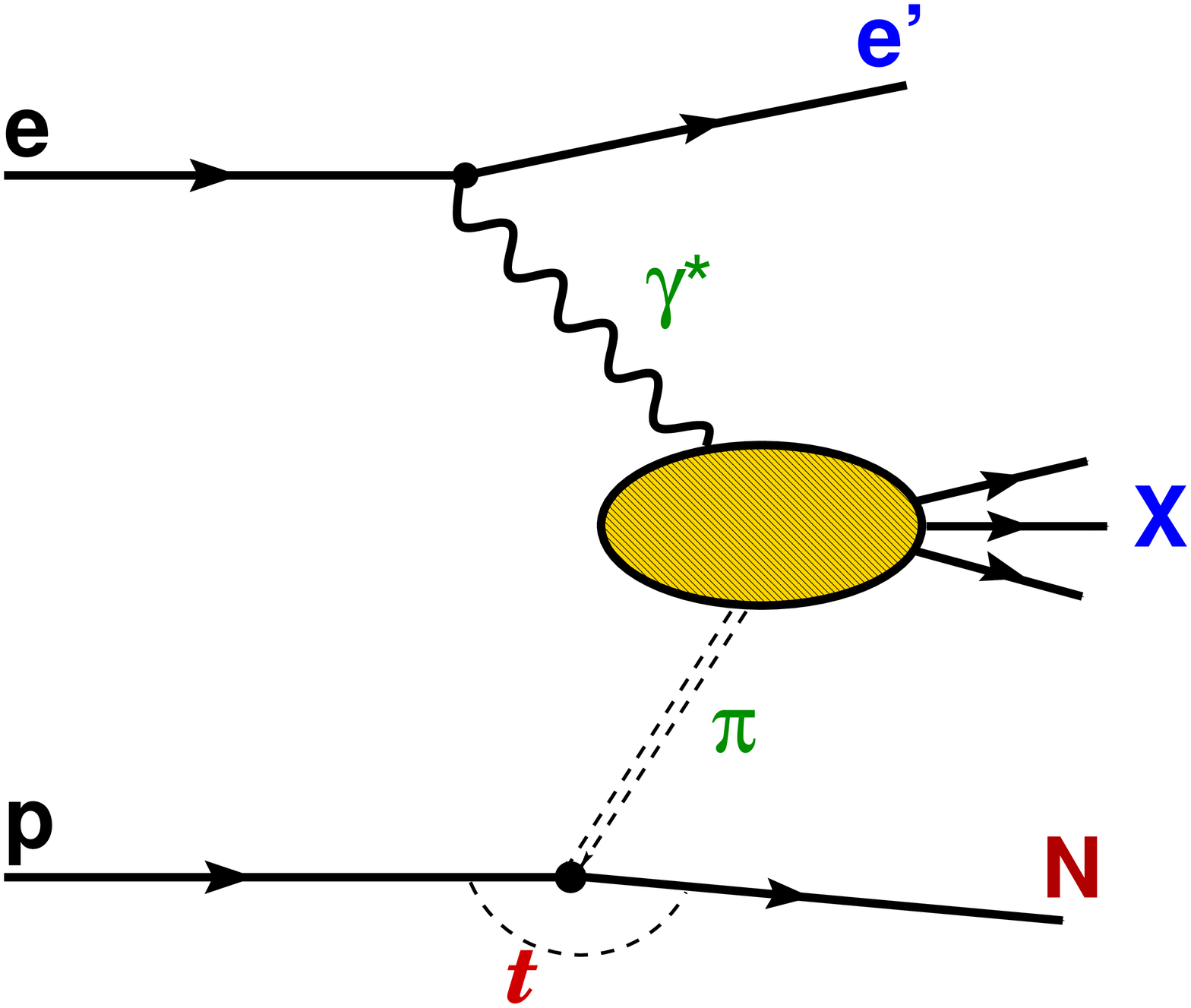,clip=,,width=6cm}
 \end{minipage} 
 \end{tabular} 
\caption{a) HERA $ep$ scattering event with the final-state baryon
in the proton-fragmentation system, $X$.
b) Leading baryon production via an exchange process.}
\label{fig-procs}
\end{figure}

In $ep$ scattering at HERA, a significant fraction of events
contains a low-transverse-momentum
baryon carrying a large fraction of the incoming
proton 
energy~\cite{pl:b384:388,*epj:c6:587,np:b637:3,np:b619:3,*epj:c41:273,np:b658:3,np:b596:3,*pl:b590:143,*pl:b610:199}.
Although the production mechanism of these
leading baryons is not completely understood,
exchange models (\fig{procs}) give a reasonable description of the data.
In this picture, the incoming proton emits a virtual particle
which scatters on the photon emitted from the beam electron.
The outgoing baryon, of energy $E_B$,
carries a fraction $x_L=E_B/E_p$ of the beam energy,
while the exchanged particle participates in
the process with energy $(1-x_L)E_p$.

In particular, one-pion exchange is a significant contributor to
leading neutron production for 
large $x_L$\,\cite{pl:b384:388,*epj:c6:587,np:b637:3}. 
For such a process the cross section for the semi-inclusive reaction
$\gamma^* p \rightarrow Xn$ factorizes into two terms
(Regge factorization~\cite{BISH}):
\begin{displaymath}
\frac{d^2 \sigma(W^2,Q^2,x_L, t)}{dx_L dt} =
        f_{\pi/p}(x_L,t) \sigma_{\gamma^* \pi}((1-x_L)W^2,Q^2) ,
\end{displaymath}
where $Q^2$ is the virtuality of the exchanged photon,
$W$ is the center-of-mass energy of the virtual photon-proton system
and $t$ is the square of the four-momentum of the exchanged pion.
In terms of the measured quantities $x_L$
and transverse momentum $p_T$, the pion virtuality is:
\begin{displaymath}
 t \simeq - \frac{p_T^2}{x_L} - \frac{(1-x_L)(m_n^2 - m_p^2 x_L)}{x_L} .
\end{displaymath}
The flux of virtual pions emitted by the proton 
is represented by $f_{\pi/p}$
and $\sigma_{\gamma^* \pi}$ is the
cross section of the virtual-photon and virtual-pion interaction
at center-of-mass energy $\sqrt{1-x_L} \, W$.
If the $ \gamma^* \pi $ cross section is independent of $t$,
the $p_T$ distribution of produced neutrons is completely determined
by the flux factor.

Many parameterizations of the pion flux have been
suggested in the
literature~\cite{BISH,FMS,*GKS,pl:b338:363,KPP,*PSI,*MST,NSSS,SNS}.
They have the general form:
\begin{displaymath}
f_{\pi/p}(x_L,t) \propto \frac{-t}{(t - m_{\pi}^2)^2}
                         (1-x_L)^{\alpha(t)} F^2(x_L,t) .
\end{displaymath}
The power $\alpha(t)$
and the form factor $F(x_L,t)$ are model dependent
with parameters that can be extracted from hadron-hadron scattering data.

Comparisons between cross sections
for the production of particles in the fragmentation region of a target
nucleon provide tests of the concepts of vertex factorization
and limiting fragmentation~\cite{pr:188:2159,*pr:d50:590}.
The hypothesis of limiting fragmentation
states that, in the high-energy limit, the production
of particles in the proton target-fragmentation region
is independent of the nature of the incident projectile.
For leading neutron production in $ep$ scattering,
where the projectile is the exchanged virtual photon,
this implies that the dependence of the cross section
on the lepton variables $(W,Q^2)$ should be independent
of the baryon variables $(x_L,t)$.
For such vertex factorization, the cross section can be written as
\begin{displaymath}
\frac{d^2 \sigma(W^2,Q^2,x_L,p_T^2)}{dx_L dp_T^2} =
        g(x_L,p_T^2) G(W^2,Q^2) ,
\end{displaymath}
where $g$ and $G$ are arbitrary functions.
The Regge factorization introduced earlier violates this vertex
factorization because of the dependence of
$\sigma_{\gamma^* \pi}$ on $x_L$ in addition to $W^2$ and $Q^2$.
Factorization tests involve comparing
semi-inclusive rates, normalized to their respective total cross sections,
to study whether particle production from a given target
is independent of the lepton variables.

In exchange models, neutron absorption can occur
through rescattering~\cite{\NSZ,\DAP,epj:c47:385,Khoze:2006hwdo}.
In a geometrical picture~\cite{\DAP},
if the size of the $n$-$\pi$ system is small compared to
the size of the photon, the neutron can also scatter
on the photon.
The neutron may migrate to lower $x_L$ and higher $p_T$
such that it is outside of the detector acceptance.
The rescattering can also transform the neutron into a charged
baryon which may also escape detection.
Since the size of the virtual photon
is inversely related to $Q^2$,
more neutron rescattering would be expected for photoproduction
($Q^2 \approx 0$)
than for deep inelastic scattering 
(DIS, $Q^2 \gtrsim 1 \gev^2$).
A previous study~\cite{np:b637:3}
showed a mild violation of vertex factorization 
with the expected increase of rate when going from photoproduction to DIS.
Similar effects have also been seen for leading
protons~\cite{np:b619:3}.
The size of the $n$-$\pi$ system is inversely proportional
to the neutron $p_T$, so rescattering removes neutrons with large $p_T$.
Thus rescattering results in a depletion of
high $p_T$ neutrons in photoproduction relative to DIS:
a violation of vertex factorization.
Pion-exchange models~\cite{BISH,FMS,*GKS,pl:b338:363,KPP,*PSI,*MST,NSSS,SNS}
incorporate a variation of the mean size of the $n$-$\pi$ system 
as a function of $x_L$.
This results in an $x_L$ dependence of the absorption,
again a violation of vertex factorization.

Absorption is a key ingredient in calculations of gap-survival
probability in $pp$ interactions at the LHC, critical in interpreting
hard diffractive processes, including central exclusive Higgs production.
The most recent absorption model
calculations~\cite{epj:c47:385,Khoze:2006hwdo},
based on multi-Pomeron exchanges,
gave a good description of
previous leading-neutron results on absorption~\cite{np:b637:3}.

This paper presents measurements of the $x_L$ and $p_T^2$ distributions
of leading neutrons
coming from samples of DIS and photoproduction $(\gamma p)$ processes,
with more than seven times higher statistics and
smaller systematic uncertainties than the previous
ZEUS publication~\cite{np:b637:3}.
The $x_L$ and $p_T^2$ distributions in DIS and photoproduction
are compared as a test of vertex factorization.
The neutron measurements are compared to similar measurements
of leading protons at HERA.
The data are also compared to the predictions of
several Monte Carlo (MC) models.
The neutron $p_T^2$ distributions in DIS are compared to several
pion-exchange models with various choices of their parameters.
Finally, the $x_L$ and $p_T$ distributions in 
photoproduction and DIS are compared to
models incorporating pion exchange and rescattering,
and a model that also includes secondary meson exchanges.

\section{Detectors}
\label{sec-det}

\Zdetdesc

\Zcaldesc\,
The EMC sections were used to detect scattered positrons in
DIS events and the RCAL was used to trigger on the dissociated
photon in photoproduction events.

Bremsstrahlung,
$ep \rightarrow e \gamma p$,
and the photoproduction of hadrons,
$ep \rightarrow e X$,
are tagged using the luminosity detectors
\cite{desy-92-066,*zfp:c63:391,*acpp:b32:2025}.
The bremsstrahlung photons are measured with a
lead-scintillator calorimeter
located at $Z=-107\met${\myZcoosysfnA} from the interaction point in the
positron-beam direction.
The positron tagger was
a similar calorimeter at
$Z=-35\met$ from the interaction point  
with an
energy resolution of $\sigma(E)/E=0.19/\sqrt{E}$ ($E$ in $\Gev$).
It was used to measure positrons scattered
at very small angles in an energy range of 5-20~GeV.

\begin{figure}[t]
\centering
\epsfig{file=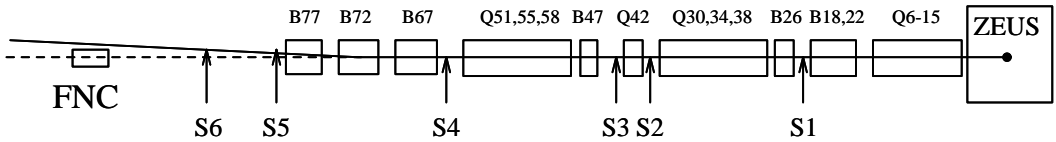,width=14cm}
\caption{ 
Side view of the proton beamline downstream from the ZEUS interaction region.
The protons are moving from right to left.
The labels for the HERA components, e.g. B47,
indicate the horizontal distance in meters from the interaction point.
The horizontal and vertical axes are not to scale. 
The proton beam is bent upward by approximately 6 mrad by the dipole
magnets B67--B77 near Z=+70 m.
The FNC is located on the zero-degree line at Z=+105.5 m.
S1--S6 indicate the locations of the ZEUS leading-proton spectrometer
stations~\pcite{zfp:c73:253}.
}
\label{fig-beamline}
\end{figure}

\begin{figure}[b]
\centering
\epsfig{file=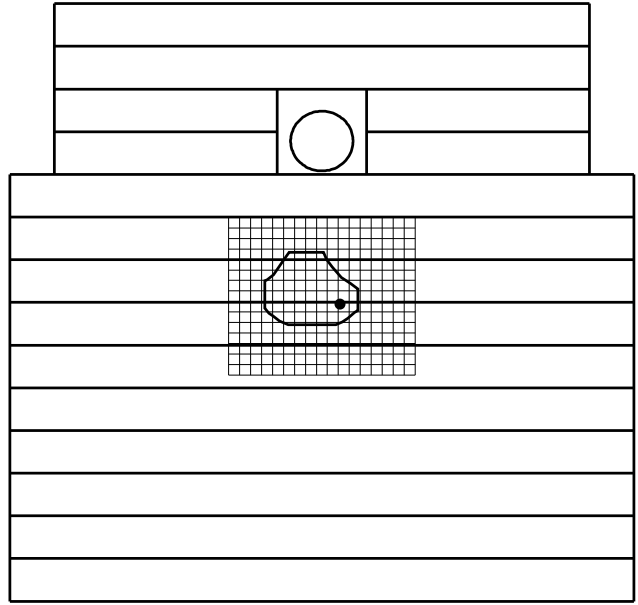,height=8cm}
\caption{
Diagram of the FNC/FNT assembly.
The thick horizontal lines show the 5 cm
vertical segmentation of the front part of the FNC.
The hole through the third and fourth towers from the top
allows the proton beam to pass through the calorimeter.
The $17 \times 15$ grid of small squares shows the fingers
of the FNT hodoscopes.
The irregular curve shows the geometric aperture defined by
upstream beamline elements,
and the bullet $(\bullet)$ shows the zero-degree point.
}
\label{fig-detector}
\end{figure}

The forward neutron calorimeter (FNC)
\cite{nim:a354:479,*nim:a394:121,*proc:calor97:295}
was installed in the HERA tunnel at 
\mbox{$\theta = 0$} degrees 
and at $Z = 106\met$ from the interaction point in the 
proton-beam direction, 
as depicted in \fig{beamline}.
It was used for the 1995-2000 data taking.
The FNC was a lead-scintillator calorimeter with an
energy resolution for hadrons measured in a test beam
to be $\sigma(E)/E = 0.70/\sqrt{E}$ ($E$ in $\Gev$).
The calorimeter was segmented vertically into 14 towers
as shown in \fig{detector}.
Three planes of veto counters were located in front of the FNC
to reject events in which a particle showered in dead material
along the beamline upstream of the FNC.

\begin{figure}[b]
\centering
\epsfig{file=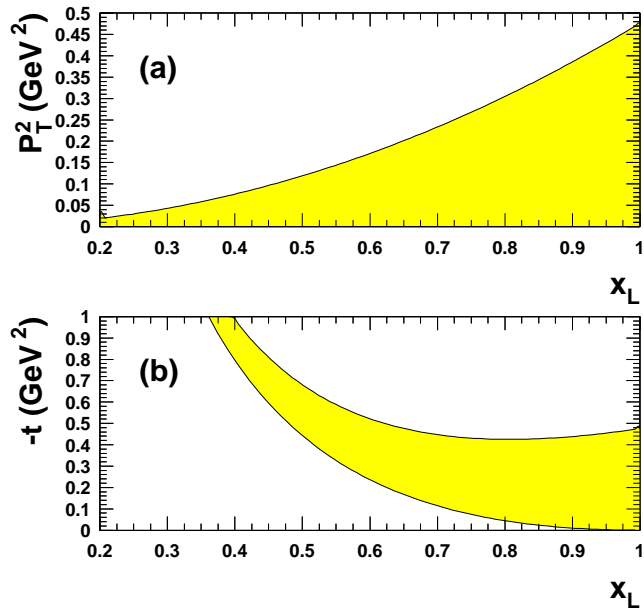,height=8cm}
\caption{
The kinematic regions in (a) $p_T^2$ and (b) $t$ covered by the
angular acceptance of the FNC ($\theta_n < 0.75 \mrad$) are
shown as shaded bands.
}
\label{fig-ranges}
\end{figure}

In 1998 a forward neutron tracker (FNT) 
was installed in the FNC at a depth of
one interaction length.
It was a scintillator hodoscope designed to
measure the position of neutron showers.
Each scintillator finger was 16.8~cm long, 1.2~cm wide and 0.5~cm deep;
17 were used for $X$ position reconstruction and 15 for $Y$.
\Fig{detector} shows the position of the FNT hodoscope
in the FNC relative to the incoming neutron beam. The irregular outlined
area indicates the geometric acceptance defined by magnet apertures.
This limited detection to neutrons with 
production angles less than \mbox{$0.75 \mrad$},
allowing transverse momenta in the range
$p_T\le E_n \, \theta_{\mbox{\scriptsize \rm max}}=0.69\, x_L \gev$.
The resulting kinematic regions in $p_T^2$ and $t$ are shown
in \fig{ranges}.

Scans by a $^{60}{\rm Co}$ radioactive source 
and data from frequent proton beam-gas runs
were used to calibrate and monitor both detectors.
The relative calibration between FNC towers was adjusted
using position information from the FNT.
The energy scale of the FNC was determined 
with a systematic uncertainty of $\pm 2\%$
from fits to the endpoint of the neutron energy spectrum
near $920 \gev$.
The minimum-ionizing-particle \mbox{(mip)}
scale in the veto counters was determined by
selecting electromagnetic showers in the FNC, a large
fraction of which were converted photons which deposited
a 2-mip signal in the counters.
The position resolution of neutron showers in the FNT of $\pm 0.23 \cm$
was measured by placing an adjustable
collimator in front of the outermost veto counter of the
FNC during special test and calibration runs.

\section{Data selection and analysis}
\label{sec-datasel}

The data for this analysis were collected in 2000 when HERA
collided $27.5\gev$ positrons with $920\gev$ protons,
giving an $ep$ center-of-mass energy $\sqrt{s}=318\gev$.
Separate triggers were used to collect DIS
and photoproduction events with leading neutrons.

\subsection{Data selection}

The DIS events were collected using a trigger that
required the detection of the
scattered positron in the CAL.
In the offline analysis,
the scattered positron was required to have energy $E_e'>10\gev$
and to be at least
3 cm from the inner edge of the beam-pipe hole in the RCAL.
The quantity $E-P_Z = \sum_i E_i (1-\cos\theta_i)$, with the sum
running over all calorimeter cells, was required to be in the
range $35<E-P_Z<65 \gev$; the lower cut reduced photoproduction
background with a misidentified positron.
These cuts resulted in a clean sample of DIS events
in the kinematic range \mbox{$Q^2 > 2\gev^2$}
with a mean photon virtuality of $\langle Q^2 \rangle \simeq 13\gev^2$.
For further studies, the variable $Q^2$ was reconstructed using the
double angle (DA) method~\cite{proc:hera:1991:23}.
This method requires a certain amount of hadronic activity
in the CAL in order to measure the angle of the hadronic system.
To ensure this, 
an additional requirement $y_{\rm JB}>0.02$
was imposed
for those measurements requiring $Q^2$ reconstruction.
Here $y_{\rm JB}$ is the inelasticity
$y \approx (W^2+Q^2)/s$ reconstructed using
the Jacquet-Blondel (JB) method~\cite{proc:epfacility:1979:391}.
The integrated luminosity of the DIS sample was approximately $40\pbi$.

The photoproduction events were collected
during the last part of the 2000 running period
using a trigger that required at least
$5\gev$ in the positron tagger in coincidence with
at least $464\mev$ in the RCAL EMC~\cite{np:b627:3}.
The acceptance of the positron tagger limited the photon
virtuality to $Q^2<0.02\gev^2$,
with a mean $Q^2$ of approximately
$\langle Q^2 \rangle \simeq 4\times10^{-4}\gev^2$.
Offline, the total energy per event deposited in the photon tagger
was required to be less than $1\gev$
in order to reject overlapping bremsstrahlung events.
The integrated luminosity of the photoproduction sample was
approximately $6\pbi$.

The DIS and photoproduction
triggers required at least $180\gev$ of energy to be deposited
in the FNC.
Good FNC neutron candidates were required offline to satisfy
the following conditions:
\begin{itemize}
\item
the FNC tower with maximum energy was one of the four towers covered by the FNT
as depicted
in \fig{detector}, to reject protons with $x_L<1$
which were deflected into the top towers of the FNC
by the vertical bending magnets shown in \fig{beamline};
\item
the veto counter had a signal of
less than one mip,
to reject showers which started in dead material upstream of
the FNC;
to minimize effects of backsplash
from hadronic showers, only the veto counter farthest
from the FNC was used;
\item
no signal in the veto counter consistent with a shower from a
previous bunch crossing, to reject pile-up energy deposits;
\item
the timing information from the FNC consistent with the triggered bunch;
\item
energy sharing among the towers was used to reject
electromagnetic showers from high-energy photons.
\end{itemize}
These cuts, similar to those used in the
previous ZEUS 
measurements~\cite{np:b637:3,np:b596:3,*pl:b590:143,*pl:b610:199},
selected long-lived neutral hadrons which had not interacted
with material before reaching the FNC.
The sample was predominantly neutrons, with a small
component of $K_L^0$ hadrons.
The MC models {\sc Rapgap} with pion
exchange~\cite{cpc:86:147}
and {\sc Lepto} with soft color interactions~\cite{pl:b366:371}
predict that the $K_L^0$ contribution is less than 2\% above
$x_L=0.6$, and increases slowly to approximately 20\% at $x_L=0.2$.

The above selection was used for the $x_L$ measurements.
For results requiring also a measurement of $p_T^2$,
the following additional requirements were imposed
to ensure a well reconstructed position measurement in the FNT:
\begin{itemize}
\item
 the highest pulse-height
channel in each of the hodoscope planes was 
above the pedestal level, to select neutrons which
showered before the FNT plane;
\item
shower profiles with more than one peak were rejected,
to eliminate mismeasurement from shower fluctuations.
\end{itemize}
The fraction of clean FNC neutrons passing the FNT cuts
determined the FNT efficiency as a function of $x_L$.
The efficiency rises with neutron energy 
from 35\% at $x_L=0.2$ to
63\% at $x_L=0.85$,
corresponding to the fraction of neutrons that shower
before the FNT.

\subsection{Normalization}
\label{sec-norm}

The cross sections for leading neutron production
presented here, $\sigma_{\rm  LN}$,
were normalized to the inclusive cross sections without
a leading-neutron requirement, $\sigma_{\rm  inc}$, as:
$$
 r_{\rm LN}(W^2,Q^2) = \frac{\sigma_{\rm LN}(W^2,Q^2)}{\sigma_{\rm inc}(W^2,Q^2)} .
$$
Variations of this relative neutron yield, $r_{\rm LN}$,
with $W^2$ or $Q^2$ indicate differences in
the neutron-production mechanism.
The acceptance for detecting different types
of events in the central ZEUS detector
in a small kinematic region of $(W^2,Q^2)$
is independent of the
neutron requirement; the acceptance cancels in the yield $r_{\rm LN}$, so that:
$$
 r_{\rm LN}(W^2,Q^2) = \frac{N_{\rm LN}(W^2,Q^2)}{N_{\rm inc}(W^2,Q^2)} .
$$
Here $N_{\rm inc}$ is the number of inclusive events in the sample
and $N_{\rm LN}$ is the number of these events with a neutron
tag, corrected for the acceptance of the forward neutron detectors.
The acceptance of the central ZEUS detector varies with $(W^2,Q^2)$;
if the neutron-tagged and inclusive events have different
kinematic dependences, their acceptances integrated over a
given $(W^2,Q^2)$ region would be different.
The mild violation of vertex factorization observed in the previous
ZEUS measurement~\cite{np:b637:3} indicates that such
differences in acceptance are less than 2\%, and so were ignored.
Thus the acceptance of the central ZEUS detector and associated
systematic uncertainties do not affect the neutron yield.
Only the acceptance of the forward neutron detectors
together with its systematic uncertainties are relevant
for measuring $r_{\rm LN}$.

For the DIS sample, a set of inclusive events
was collected simultaneously
and used to normalize the neutron data.
For the inclusive photoproduction events,
$49\nbi$ of data were collected in a special run in 1996
when the proton beam energy was $820\gev$,
for measurement of the photon-proton total cross section~\cite{np:b627:3}.
The normalization of the 2000 photoproduction data was determined
by assuming that the ratio of the photoproduction and
DIS neutron yields,
for any given neutron $(x_L,p_T^2)$ kinematic region,
was the same at the two proton beam energies:
$$
\left. \frac{r_{\rm LN}^{\gamma p}}{r_{\rm LN}^{\rm DIS}} \right|_{\rm 920\gev} =
\left. \frac{r_{\rm LN}^{\gamma p}}{r_{\rm LN}^{\rm DIS}} \right|_{\rm 820\gev} .
$$
In this equation
$r_{\rm LN}^{\gamma p}({\rm 820\gev})$ was measured in the
1996 inclusive photoproduction sample,
$r_{\rm LN}^{\rm DIS}({\rm 820\gev})$ was measured
in inclusive DIS data from the same running period
and $r_{\rm LN}^{\rm DIS}({\rm 920\gev})$ was measured
in the inclusive DIS data from the 2000 running period.
The neutron kinematic region for the measured yields was
$x_L>0.2$ and $p_T^2<0.476 \,x_L^2 \gev^2$.
This normalization procedure resulted in an
uncertainty on the neutron yield in photoproduction
of 5.1\%, predominantly from the limited statistics of the
1996 photoproduction data.

\subsection{Beamline and forward-detector simulation}

The acceptance of the forward neutron detectors,
for the measurement of neutron yields, was determined
from a simple one-particle MC simulation.
The simulation accounts for the aperture and dead material
along the neutron flight path,
the measured proton beam position and  $p_T$ spread,
and the measured detector resolutions.

\begin{figure}[p]
\centering
\epsfig{file=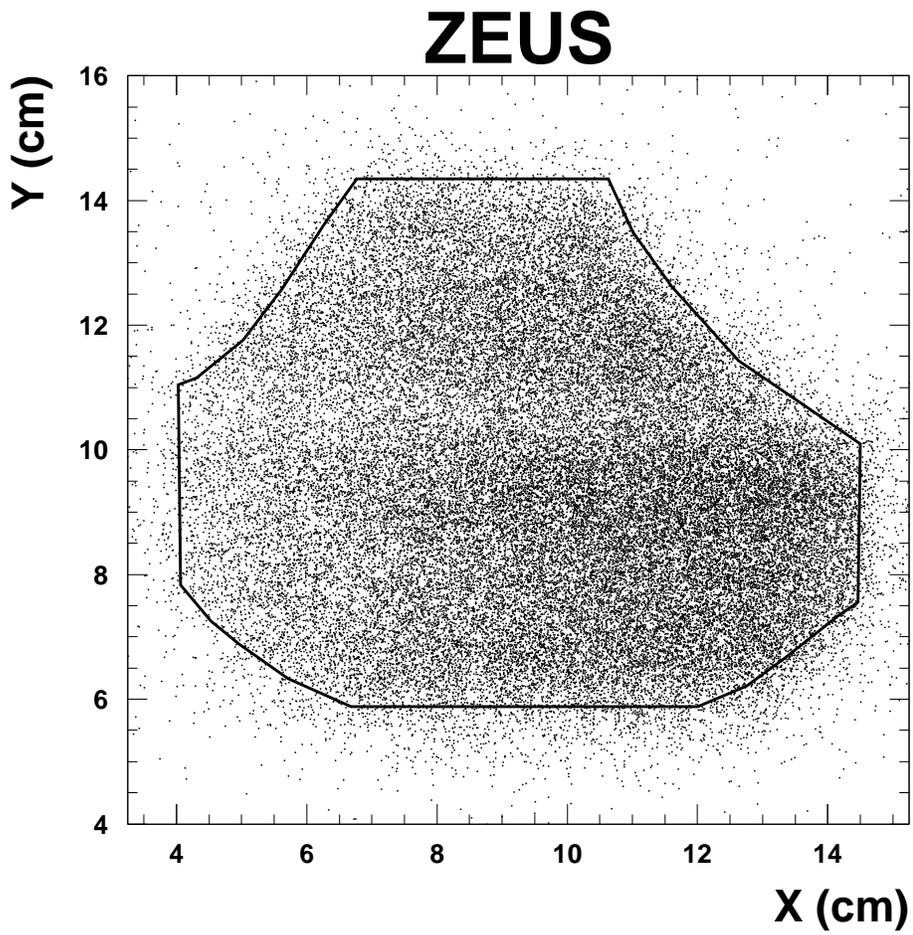,width=12cm}
\caption{
Scatter plot of reconstructed positions
in the FNT.
The irregular curve is the geometric aperture defined by
upstream beamline elements as modeled in the Monte Carlo.
The zero-degree point is at $X=12.5 \cm$, $Y=8.3 \cm$.
The deficit of events observed near $Y \sim 10.5 \cm$
is due to dead material in the LPS.
}
\label{fig-scat}
\end{figure}

\Fig{scat} shows a scatter plot of reconstructed hits in the FNT
from a sample of DIS events.
The irregular curve is the aperture expected from the MC simulation.
Numerous events are reconstructed outside of the aperture,
as the aperture is not a sharp
boundary as modeled,
but presents a varying amount of dead material
over several millimeters transverse to the neutron flight path.
The effect of this on the measured neutron yield is less
than 2\% and was ignored.

\begin{figure}[p]
\centering
\epsfig{file=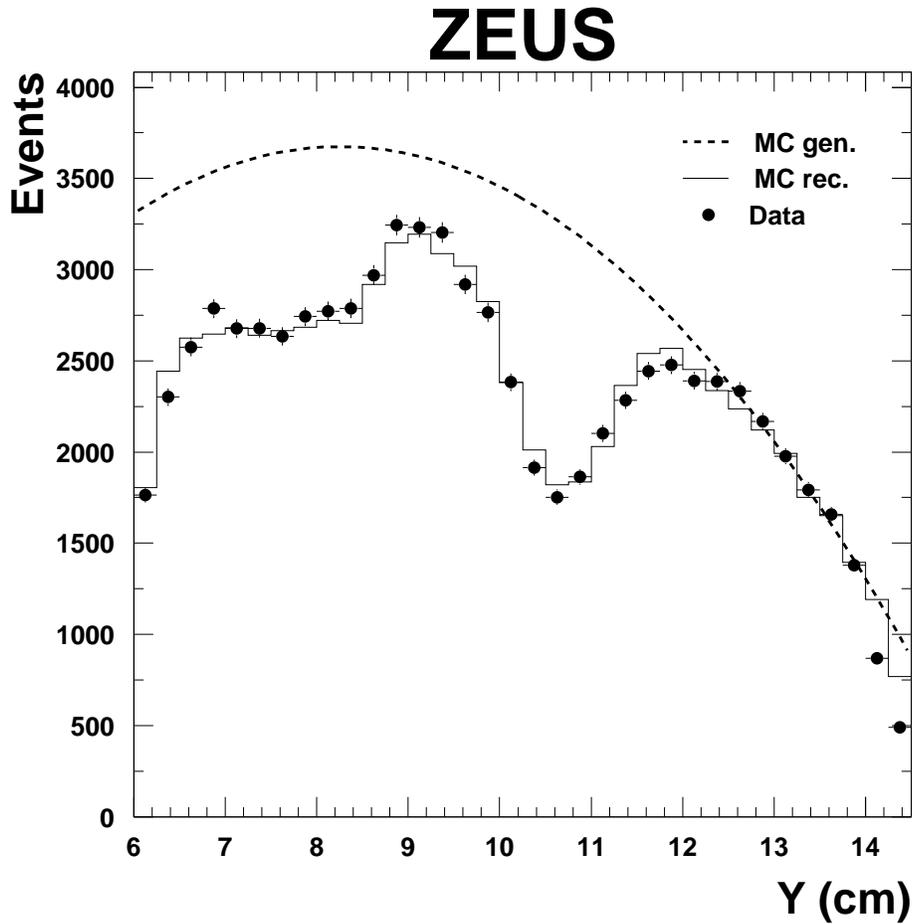,width=12cm}
\caption{
Vertical slice of the data in \fig{scat} for $7<X<10.5 \cm$.
The smooth curve is the generated Monte Carlo distribution;
the histogram is the Monte Carlo distribution
of reconstructed events after the dead material simulation.
The peak of the smooth curve,
determined from a fit to the data,
corresponds to the
zero-degree point in the vertical plane.
}
\label{fig-slice}
\end{figure}

The simulation also modeled significant amounts of dead material
along the neutron flight path,
primarily from stations S5 and S6 of the ZEUS leading
proton spectrometer (LPS) shown in \fig{beamline}.
The LPS was
a set of Roman pots used to measure protons scattered
at very low angles~\cite{zfp:c73:253}.
The elements of these stations were measured after the
LPS was removed from the HERA tunnel in 2000 and
implemented in the simulation.
The positions of these elements in the simulation
were adjusted to reproduce the data.
For example, the deficit of events observed near
$Y \sim 10.5 \cm$,
seen clearly in the vertical slice of the scatter plot in \fig{slice},
determined the vertical alignment of the LPS.
There is good agreement between the simulation and the data
distributions.

During operation, the LPS was normally in one of two positions:
extracted,
or inserted for data taking.
Separate dead material maps were made for the two positions.
Data collected during periods when the LPS was moving were rejected.
The results were determined separately for the two LPS positions and
combined according to the luminosity taken in each position.
The difference with the results obtained for each position was taken as a
measure of the systematic uncertainty from the dead material map.

The reconstruction of the neutron scattering angle, $\theta_n$, requires
knowledge of the zero-degree point.
This was determined 
by generating a symmetric distribution of neutrons,
passing it through
the simulation, and fitting the reconstructed distribution to the data.
An example of such a fit for the zero-degree position in the vertical
plane
is shown in \fig{slice}.
Considering different input distributions, and
taking into account the
uncertainties in the dead material map,
the beam zero-degree point
was determined to an accuracy of \mbox{$\pm 0.2 \cm$} in both $X$ and $Y$.

The simulation also takes into account
the energy resolution of the FNC, the position resolution of the FNT,
and the inherent $p_T$-spread of the HERA proton beam.
The latter was measured in the reaction
$\gamma p \rightarrow \rho p$, with the $\rho$ decay products
measured in the ZEUS central detector and the final-state proton
measured with the LPS~\cite{zfp:c73:253}.
The beam $p_T$-spread corresponds to
a smearing of the zero-degree point by
0.45 cm  horizontally and \mbox{1.0 cm} vertically,
significantly larger than the FNT resolution.
The spectrum of generated neutrons
was tuned to match the $x_L$ and $p_T^2$ distributions
separately for the DIS
and photoproduction samples.

These MC distributions were then used to correct
the data for all acceptance and smearing effects.
For distributions requiring a position measurement,
the correction for
the $x_L$-dependent FNT efficiency was also applied.

\subsection{Systematic uncertainties}
\label{sec-systemat}

The dominant effects contributing to the systematic uncertainties 
arose from:
\begin{itemize}
\item the beam zero-degree point;
\item the dead material map;
\item the FNC energy scale;
\item the $p_T^2$ distributions, input for $x_L$-distribution acceptance.
\end{itemize}
The systematic uncertainties were typically
5--10\% of the measured quantities, and are
shown as shaded bands in the figures.
The variation of the energy scale shifted the points in $x_L$.
The other systematic variations amount to a change in acceptance
resulting in a correlated shift of the neutron yields. 
The small uncertainties from the
assumption of acceptance cancellation in $r_{\rm LN}$
and from the aperture edge were ignored.

Corrections for 
efficiency of
the cuts and backgrounds  were applied to the
normalization of the neutron yields.
The corrections, 
similar to those of the previous
ZEUS measurement~\cite{np:b637:3}, were:
\begin{itemize}
\item false veto counter firing: $+10.6 \pm 1.0 \%$, determined from
      randomly triggered events;
\item veto counter inefficiency: $-2.5 \pm 1.0 \%$, determined from
      2-mip distributions in the veto counter;
\item backsplash from neutron showers: $+1.5 \pm 1.5 \%$, determined from
      timing information in the veto counter giving the
      fraction of late time vetoes;
\item neutrons from proton beam-gas interactions: $-1.4 \pm 0.3 \%$, 
       determined from randomly triggered events.
\end{itemize}
The overall systematic uncertainty on the normalization of the
neutron yield from these corrections,
not included in the shaded bands of the figures,
 was $ \pm 2.1 \%$.
The yield in photoproduction has an additional uncertainty of
$ \pm 5.1 \%$ from the normalization procedure described
in \Sect{norm}.


\section{Results}

\subsection{Neutron $\mathbf{x_L}$ and $\mathbf{p_T^2}$ distributions in DIS}
\label{sec-res}

\begin{figure}[p]
\centering
\epsfig{file=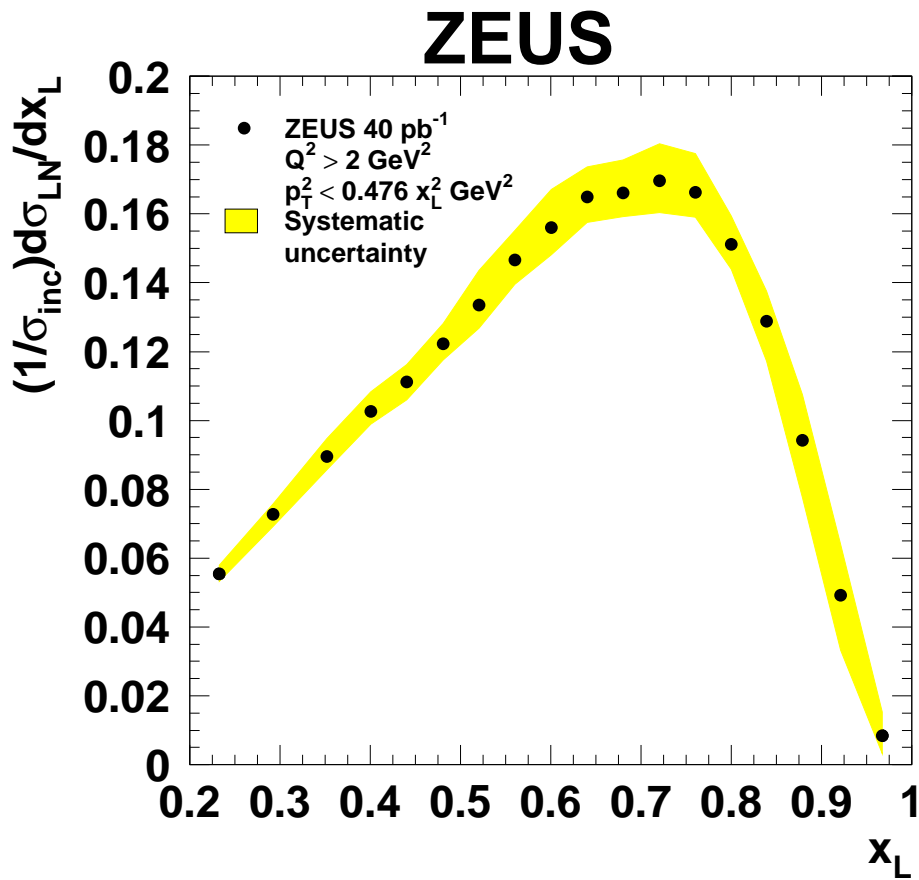,width=12cm}
\caption{
The distribution 
$ (1/\sigma_{\rm inc}) d\sigma_{\rm LN}/dx_L$
for neutrons in DIS with scattering angles $\theta_n < 0.75 \mrad$,
corresponding to the kinematic range
$p_T^2 < 0.476 \, x_L^2 \gev^2$.
The statistical uncertainties are smaller than the plotted symbols;
the shaded band shows the systematic uncertainties.
The band does not include the overall normalization uncertainty of $2.1\%$.
}
\label{fig-xlthcut}
\end{figure}

\Fig{xlthcut} shows the normalized differential distribution
$ (1/\sigma_{\rm inc}) d\sigma_{\rm LN}/dx_L $
for neutrons in DIS with scattering angles $\theta_n < 0.75 \mrad$,
corresponding to the kinematic range
$p_T^2 < 0.476 \, x_L^2 \gev^2$.
It rises from the lowest $x_L$ due to the increase in $p_T^2$ space,
reaches a maximum near $x_L=0.7$,
and falls to zero at the endpoint $x_L=1$.
These results are consistent with the previous
ZEUS measurement~\cite{np:b637:3}.
Integrating this distribution, the total leading-neutron
yield for the measured region is:
$$r_{\rm LN}(Q^2>2\gev^2,x_L>0.2,p_T^2<0.476 \, x_L^2 \gev^2)=
  0.0885 \pm 0.0002 \, {\rm (stat.)} \, ^{+0.0034}_{-0.0029} \, {\rm
  (sys.)}. $$
Here the systematic uncertainty includes the overall $\pm 2.1\%$
scale uncertainty.

\begin{figure}[p]
\centering
\epsfig{file=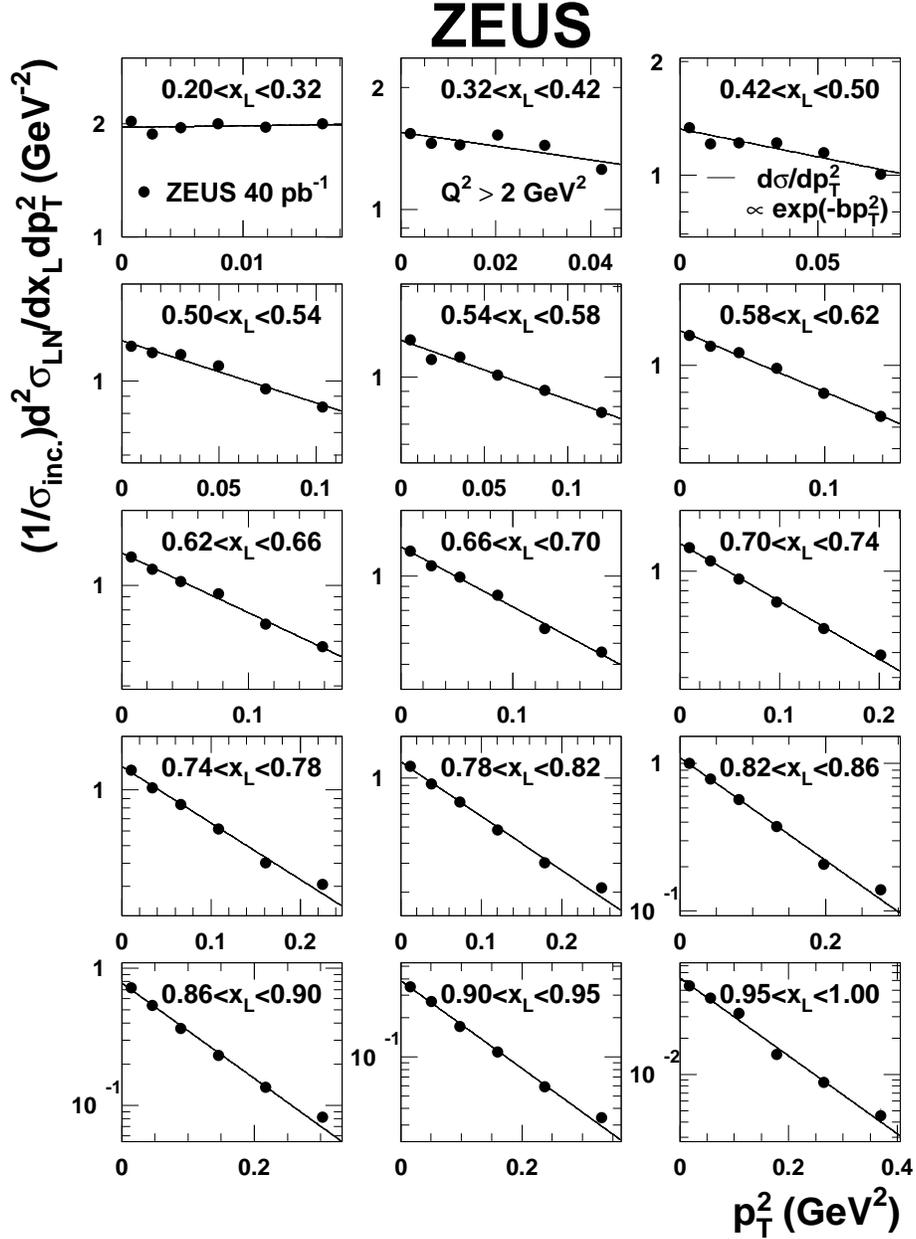,width=12cm}
\caption{
The $p_T^2$ distributions for DIS.
Note the logarithmic vertical scale and the varying $p_T^2$ ranges.
The statistical uncertainties are shown by a vertical error bar;
in most cases they are smaller than the plotted symbol. 
The systematic uncertainties
are not shown.
The line on each plot is the result of a fit to the
form \mbox{$d\sigma_{\rm LN}/dp_T^2 \propto \exp(-bp_T^2)$}.
}
\label{fig-ptsq_dis}
\end{figure}

\begin{figure}[p]
\centering
\epsfig{file=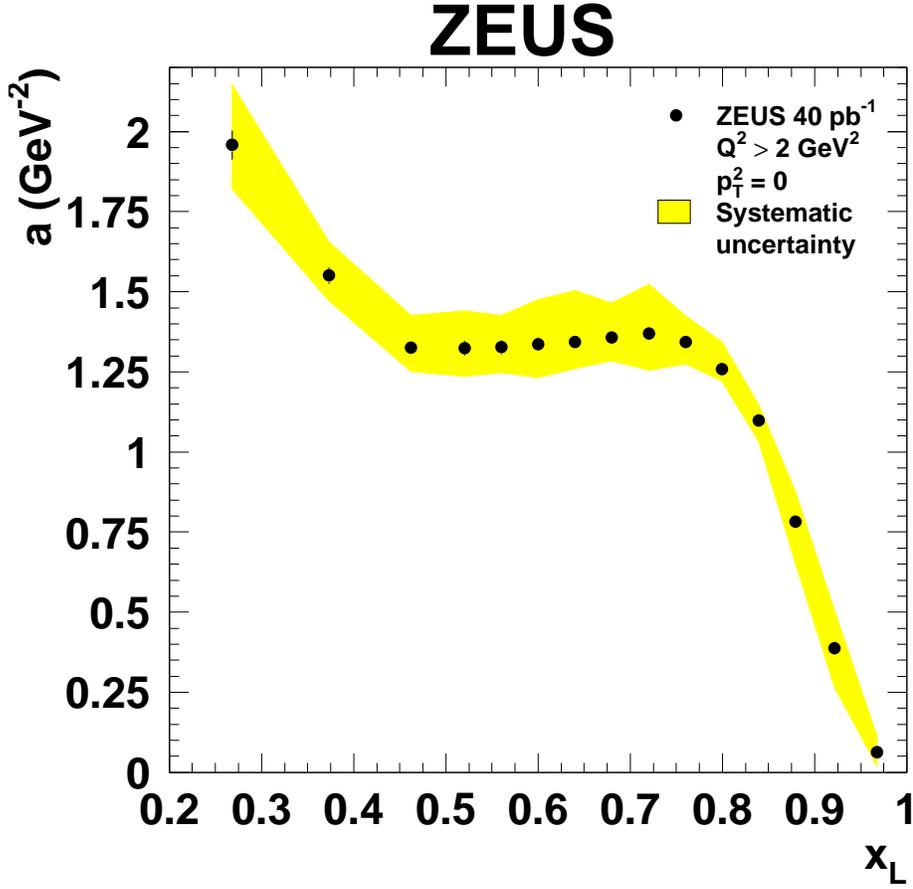,width=12cm}
\caption{
The intercepts
$ a = \left (1/\sigma_{\rm inc}) d^2 \sigma_{\rm LN}/dx_L dp_T^2 \right|_{p_T^2=0}$
versus $x_L$ from fits of the $p_T^2$ distributions
in DIS 
to the form $d\sigma_{\rm LN}/dp_T^2 = a \exp(-bp_T^2)$
over the $p_T^2$ ranges shown in \fig{ptsq_dis}.
The statistical uncertainties are shown by a vertical error bar;
in most cases they are smaller than the plotted symbol. 
The shaded band shows the systematic uncertainties.
The band does not include the overall normalization uncertainty of $2.1\%$.
}
\label{fig-intxlsys}
\end{figure}

\begin{figure}[p]
\centering
\epsfig{file=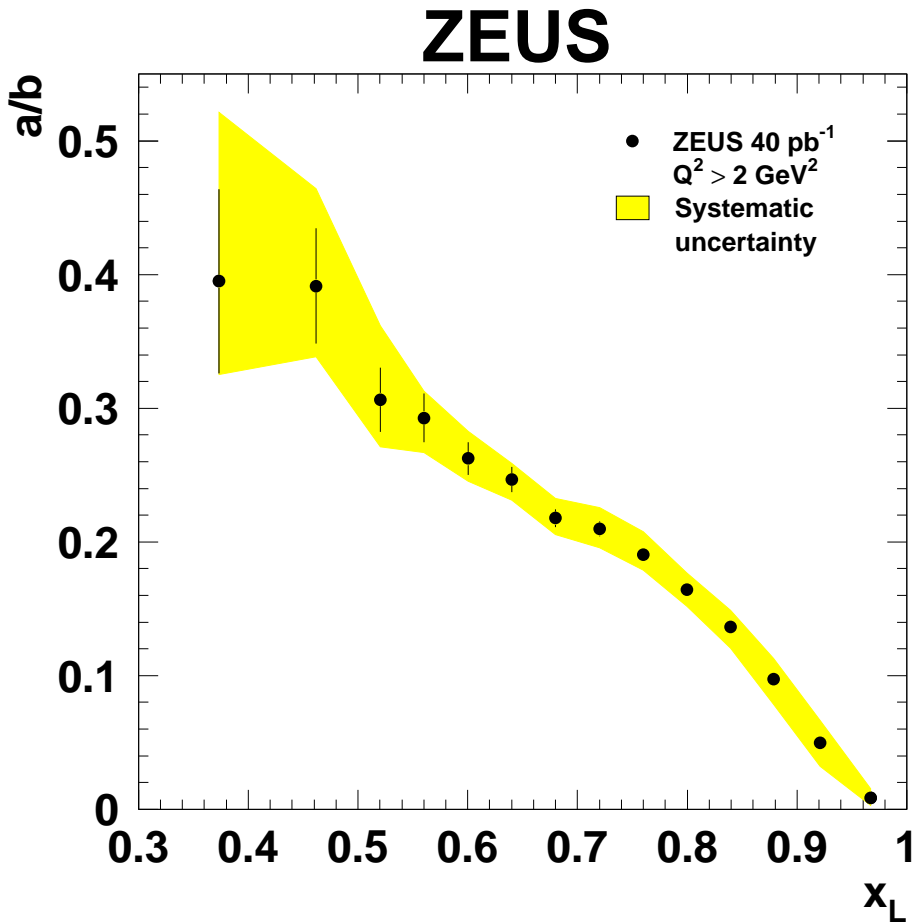,width=12cm}
\caption{
The ratio $a/b$
versus $x_L$ from fits of the $p_T^2$ distributions
in DIS 
to the form $d\sigma_{\rm LN}/dp_T^2 = a \exp(-bp_T^2)$
over the $p_T^2$ ranges shown in \fig{ptsq_dis}.
The error bars show the statistical uncertainties;
the shaded band shows the systematic uncertainties.
The band does not include the overall normalization uncertainty of $2.1\%$.
}
\label{fig-aobsys}
\end{figure}

The corrected $p_T^2$ distributions in DIS for different
$x_L$ bins are shown in
\fig{ptsq_dis} and summarized in \tab{dndpdxldis}.
They are presented as normalized doubly differential distributions
$ (1/\sigma_{\rm inc}) d^2 \sigma_{\rm LN}/dx_L dp_T^2$.
The bins in $p_T^2$ are at least as large as the
resolution, which is dominated by the $p_T$ spread of the proton beam.
The varying $p_T^2$ ranges of the plots are due
to the aperture limitation.
The line on each plot is a fit to the functional form
$d\sigma_{\rm LN}/dp_T^2 \propto \exp(-bp_T^2)$.
The distributions are compatible with a single
exponential within the statistical and uncorrelated systematic uncertainties.
Thus, with the parameterization
$$
 \frac{1}{\sigma_{\rm inc}} \frac{d^2 \sigma_{\rm LN}}{dx_L dp_T^2} =
               a(x_L) \:  e^{\displaystyle -b(x_L) p_T^2} ,
$$
the neutron $(x_L,p_T^2)$ distribution is fully
characterized by the slopes $b(x_L)$ and intercepts
$ a(x_L) =
\left. (1/\sigma_{\rm inc}) d^2 \sigma_{\rm LN}/dx_L
dp_T^2\right|_{p_T^2=0} $.
The dependences of the intercepts
and the slopes on $x_L$ are shown in
\figand{intxlsys}{bxlsys} and summarized in \tab{abdis}.
The systematic uncertainties were evaluated by
making the variations listed in \Sect{systemat}
and repeating the fits.
The intercepts fall rapidly from the lowest $x_L$,
are roughly constant in the region $x_L=0.45$-$0.75$,
and fall to zero at the endpoint $x_L=1$.
Below $x_L=0.32$, the slopes are consistent with zero and are not plotted;
they rise linearly in the range $0.3<x_L<0.85$ to a value of
$b \simeq 8 \gev^{-2}$, and then decrease slightly at higher $x_L$.
\Fig{aobsys} shows the ratio $a/b$ for the region 
$0.32<x_L<1$ where $b>0$, 
which can be taken as the leading-neutron
yield integrated over $p_T^2$ values from zero to infinity,
 assuming that the
$p_T^2$ distributions remain
an exponential 
also beyond the
measured $p_T^2$ range. This distribution, integrated over $x_L$ in the range
$0.32<x_L<1$, corresponds to a yield:
$$r_{\rm LN}(Q^2>2\gev^2,x_L>0.32)=
  0.159 \pm 0.008 \, {\rm (stat.)} \, ^{+0.019}_{-0.006} \, {\rm (sys.)}. $$
Here the systematic uncertainty includes the overall 2.1\%
scale uncertainty.

\subsection{$\mathbf{Q^2}$ dependence of leading neutron production}
\label{sec-lnqqdep}

To investigate the $Q^2$ dependence of leading-neutron production,
the full DIS sample
was divided into three subsamples depending on the
$Q_{\rm DA}^2$ range, with the additional cut $y_{\rm JB}>0.02$.
The kinematic regions in $Q^2$ and $W$ for all DIS and photoproduction
samples are summarized in \tab{kinranges}.

\begin{figure}[p]
\centering
\epsfig{file=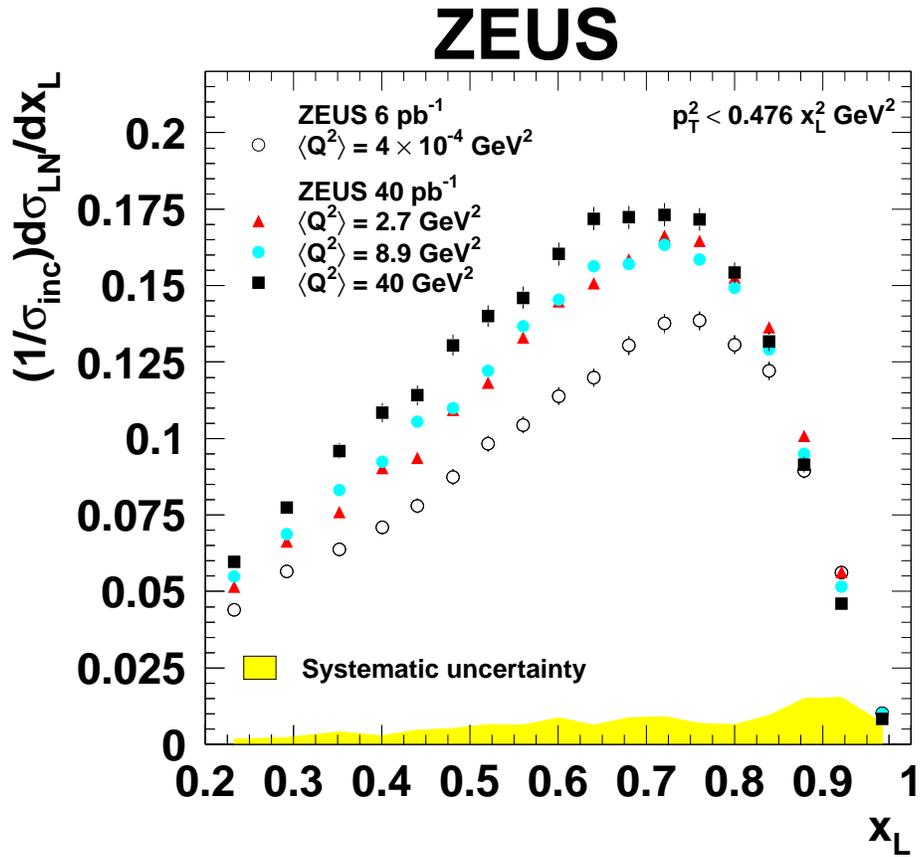,width=12cm}
\caption{
The $x_L$ distributions for the photoproduction and three DIS subsamples.
The error bars show the statistical uncertainties.
The common systematic uncertainties are shown as a shaded band.
There is an overall normalization uncertainty of $2.1\%$
for the DIS data, and an additional uncorrelated uncertainty
of $5.1\%$ for the photoproduction data.
}
\label{fig-xlsqq}
\end{figure}

\begin{figure}[p]
\centering
\epsfig{file=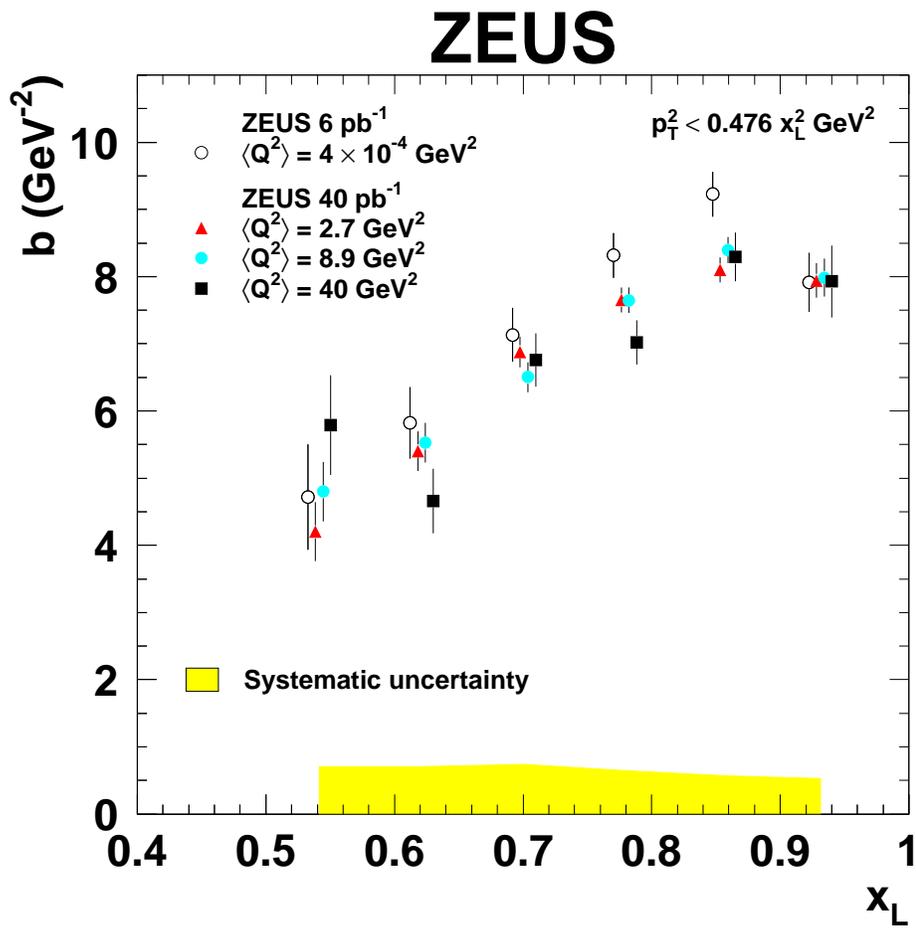,width=12cm}
\caption{
The exponential slopes $b$ for the photoproduction and three DIS subsamples.
The points are slightly offset horizontally for clarity.
The error bars show the statistical uncertainties.
The common systematic uncertainties are shown as a shaded band.
}
\label{fig-bqq}
\end{figure}

\Fig{xlsqq} shows the $x_L$ distributions and \fig{bqq} the $p_T^2$ slopes
in the range $p_T^2 < 0.476 \, x_L^2 \gev^2$ for the photoproduction and three
DIS subsamples.
There is a trend of increasing neutron yield with increasing $Q^2$,
a clear violation of vertex factorization.
There is a large increase between the photoproduction region
and the low- and mid-$Q^2$ DIS regions, 
in which the data are nearly $Q^2$ independent.
There is then a smaller, but significant, increase between the mid-$Q^2$
and high-$Q^2$ regions.
The effect of the different $W$ ranges for the DIS and
photoproduction samples on the neutron yield was less than 5\%,
as evaluated by restricting the DIS sample to low- and high-$W$ regions.
The slopes for all three DIS samples are equal within the uncertainties.
The slopes for photoproduction are higher in the region
$0.6<x_L<0.9$.

The total neutron yields integrated over $x_L>0.2$ for the four
samples are summarized in \tab{intxl}.
The doubly differential distributions
$ 1/\sigma_{\rm inc} \: d\sigma_{LN}/dx_L dp_T^2$
for the photoproduction and three DIS subsamples are summarized
in \tab{dndpdxlqq},
and the intercepts and slopes of the exponential fits
are summarized in \tab{abqq}.

To investigate the differences between the photoproduction
and DIS regimes further, 
the effects of energy calibration and beam position drifts
were minimized by using
only the subset of DIS data collected simultaneously with
the photoproduction data.
The DIS sample without $y_{\rm JB}$ or $Q_{\rm DA}^2$ cuts
was used to maximize the statistical precision of the comparison.

\begin{figure}[p]
\centering
\epsfig{file=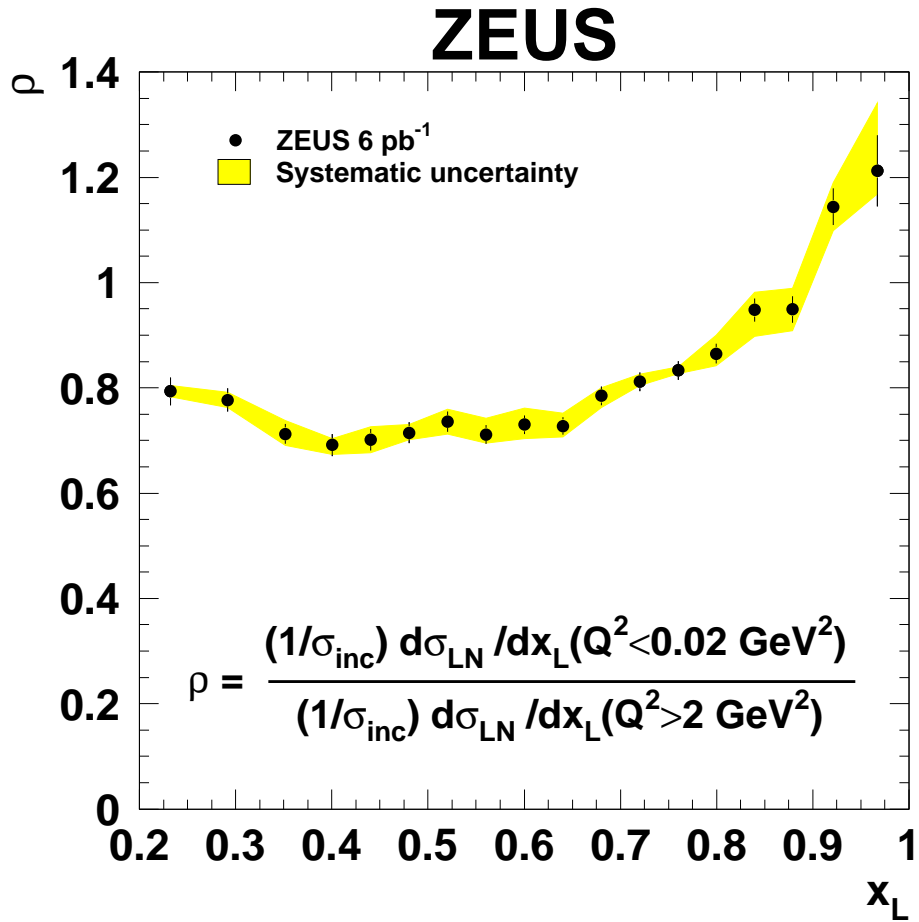,width=12cm}
\caption{
The ratio of the $x_L$ distributions
for photoproduction and DIS,
normalized as described in the text,
for neutrons with scattering angles $\theta_n < 0.75 \mrad$,
corresponding to the kinematic range
$p_T^2 < 0.476 \, x_L^2 \gev^2$.
The error bars show the statistical uncertainties;
the shaded band shows the systematic uncertainties. 
The band does not include the overall normalization
uncertainty of $5.1\%$ on the ratio.
}
\label{fig-xlphpdis}
\end{figure}

The ratio of the normalized differential distributions
\begin{displaymath}
 \rho(x_L) = \frac{
\frac{\textstyle 1}{\textstyle \sigma^{\gamma p}_{\rm inc}} \,
   \frac{\textstyle d \sigma^{\gamma p}_{\rm LN}}{\textstyle d x_L}
                }{
\frac{\textstyle 1}{\textstyle \sigma^{\rm DIS}_{\rm inc}} \,
   \frac{\textstyle d \sigma^{\rm DIS}_{\rm LN}}{\textstyle d x_L}
                }
\label{eqn-norm}
\end{displaymath}
for the region $p_T^2 < 0.476 \, x_L^2 \gev^2$
is shown in \fig{xlphpdis}.
In the range \mbox{$0.2<x_L<0.4$}, the ratio drops slightly
but rises for higher $x_L$ values, exceeding unity for $x_L>0.9$.
The deviation of the ratio from
unity is a clear violation of vertex factorization.
The ratio of the intercepts for photoproduction and DIS,
which has a nearly identical behavior to that of
\fig{xlphpdis}, is not shown.

\begin{figure}[p]
\centering
\epsfig{file=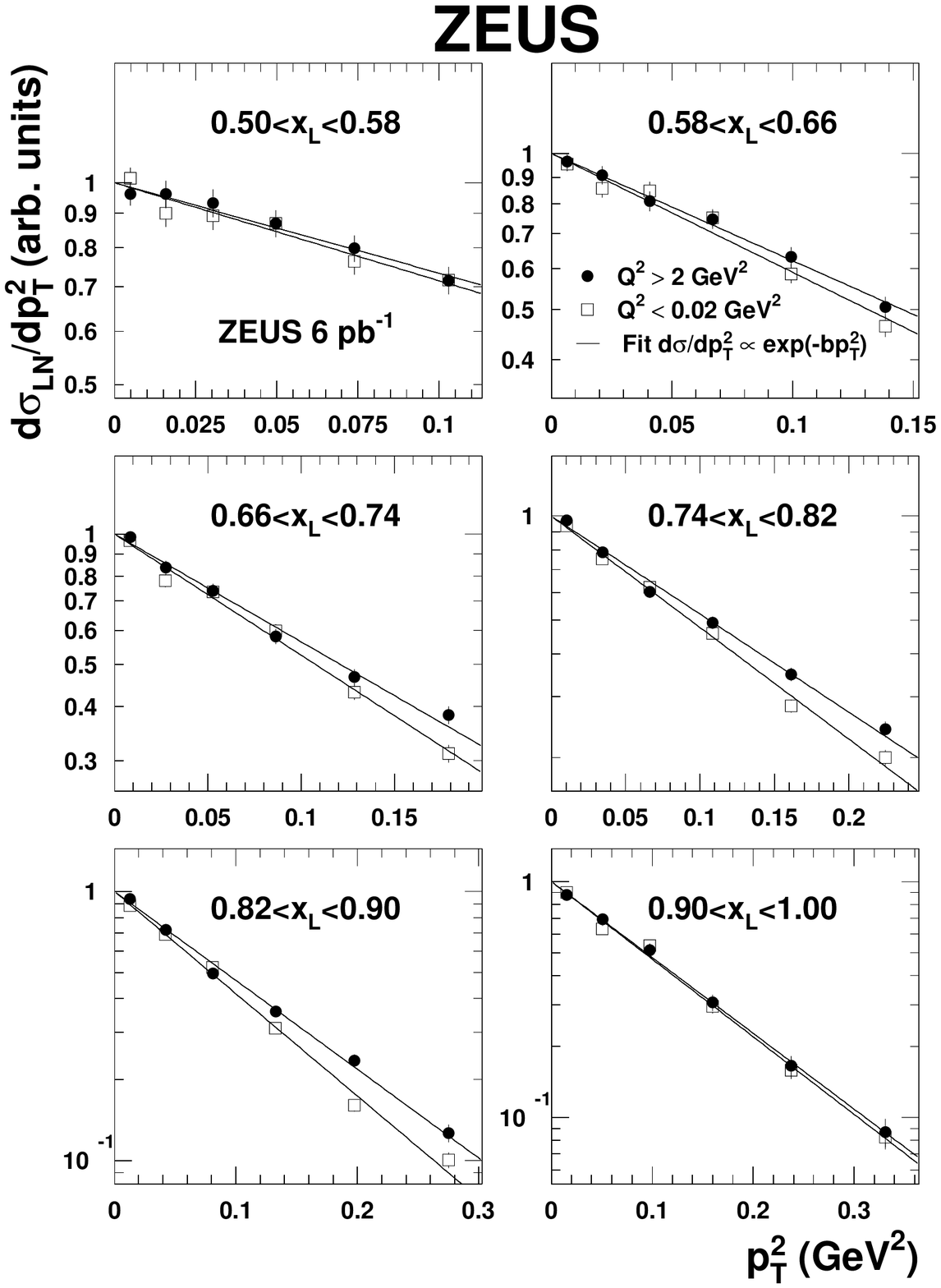,width=12cm}
\caption{
The $p_T^2$ distributions for photoproduction and DIS.
They are each normalized to unity at $p_T^2 = 0$.
Note the logarithmic vertical scale and the varying $p_T^2$ ranges.
The statistical uncertainties are shown by a vertical error bar;
in some cases they are smaller than the plotted symbol.
The systematic uncertainties
are not shown.
The lines on each plot are the results of fits to the
form \mbox{$d\sigma_{\rm LN}/dp_T^2 \propto \exp(-bp_T^2)$}.
}
\label{fig-ptsq_phpdis}
\end{figure}

\begin{figure}[p]
\centering
\epsfig{file=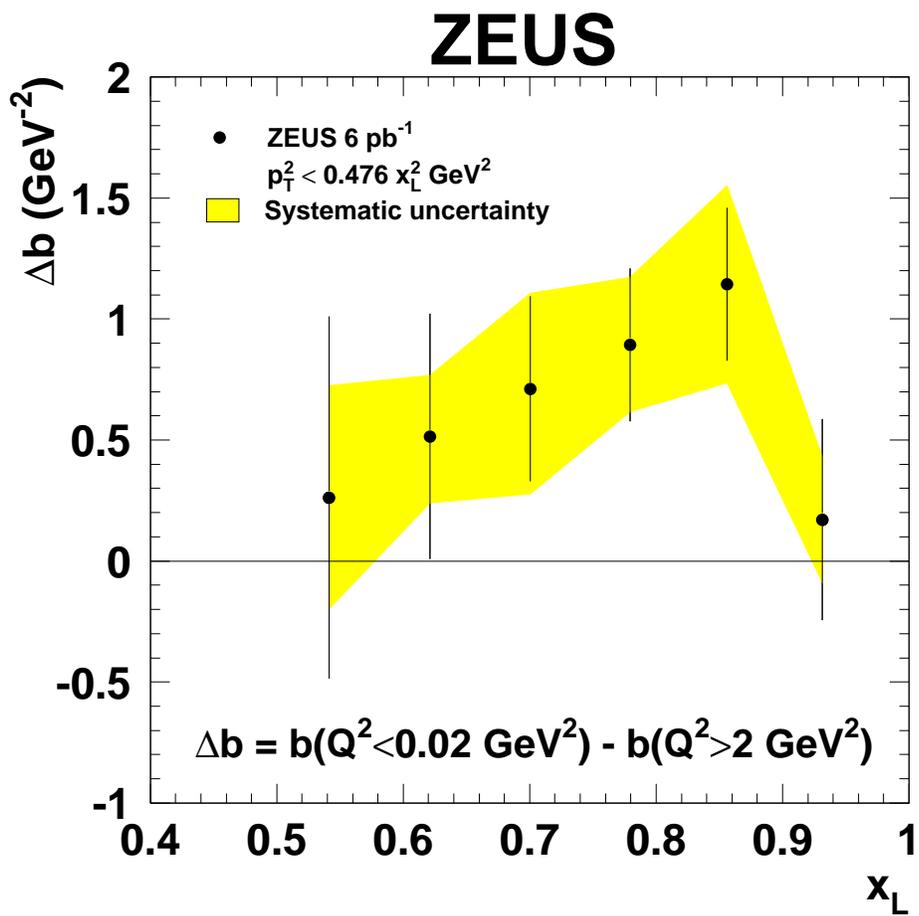,width=12cm}
\caption{
The differences between the exponential
slopes in photoproduction and DIS, $\Delta b$, versus $x_L$.
The error bars show the statistical uncertainties;
the shaded band shows the systematic uncertainties.
}
\label{fig-dbxlsys}
\end{figure}

The $p_T^2$ distributions for both samples, normalized
to unity at $p_T^2=0$, are shown in \fig{ptsq_phpdis}.
The photoproduction distributions are 
steeper in the range
$0.6<x_L<0.9$, with relatively fewer neutrons at high $p_T^2$.
The difference of the slopes
$\Delta b = b(Q^2<0.02 \gev^2)-b(Q^2>2 \gev^2)$
is less sensitive to systematic effects than each of
the individual slopes.
These values are plotted in \fig{dbxlsys}.
The slopes for photoproduction are 
larger in the range $0.6<x_L<0.9$,
with \mbox{$\Delta b = $ 0.5--1.0 $\gev^{-2}$,}
qualitatively consistent with the violation of vertex factorization
expected from absorption as discussed in the introduction.

\subsection{Comparison to leading protons}
\label{sec-rescomp}

\begin{figure}[p]
\centering
\epsfig{file=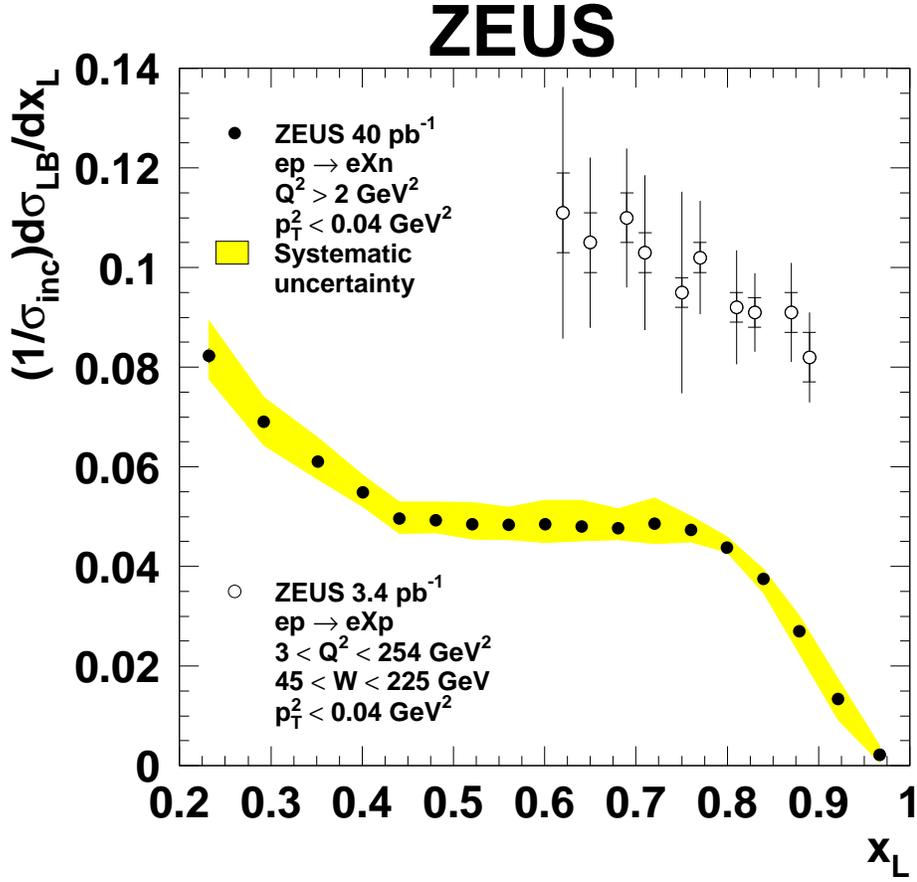,width=12cm}
\caption{
The normalized differential distributions
$ (1/\sigma_{\rm inc}) d\sigma_{\rm LB}/dx_L$ 
for leading protons and neutrons in the range
$p_T^2<0.04 \gev^2$.
For the leading-neutron points, 
the statistical uncertainties are shown by a vertical error bar;
in most cases they are smaller than the plotted symbol.
The shaded band shows the systematic uncertainties;
the band does not include the overall normalization uncertainty of $2.1\%$.
For the leading-proton points, the inner error bars show
the statistical uncertainty only;
the full error bars show the statistical and systematic
uncertainties added in quadrature.
}
\label{fig-lpsxl}
\end{figure}

ZEUS has also reported $x_L$ 
distributions for leading
protons in the kinematic region
\mbox{$p_T^2<0.04 \gev^2$}~\cite{np:b658:3}.
The neutron $x_L$ distributions were also measured in the same
region, using the FNT measurement of $p_T^2$.
The results for DIS are compared in \fig{lpsxl}.
There are approximately twice as many protons as neutrons in the range
$0.6<x_L<0.9$. 
If only isospin-1 particle exchanges contributed to proton production,
there should be half as many protons as neutrons.
Thus, exchanges of particles with different isospins
such as isoscalars
must be invoked to account for the observed proton rate~\cite{SNS}.

\begin{figure}[p]
\centering
\epsfig{file=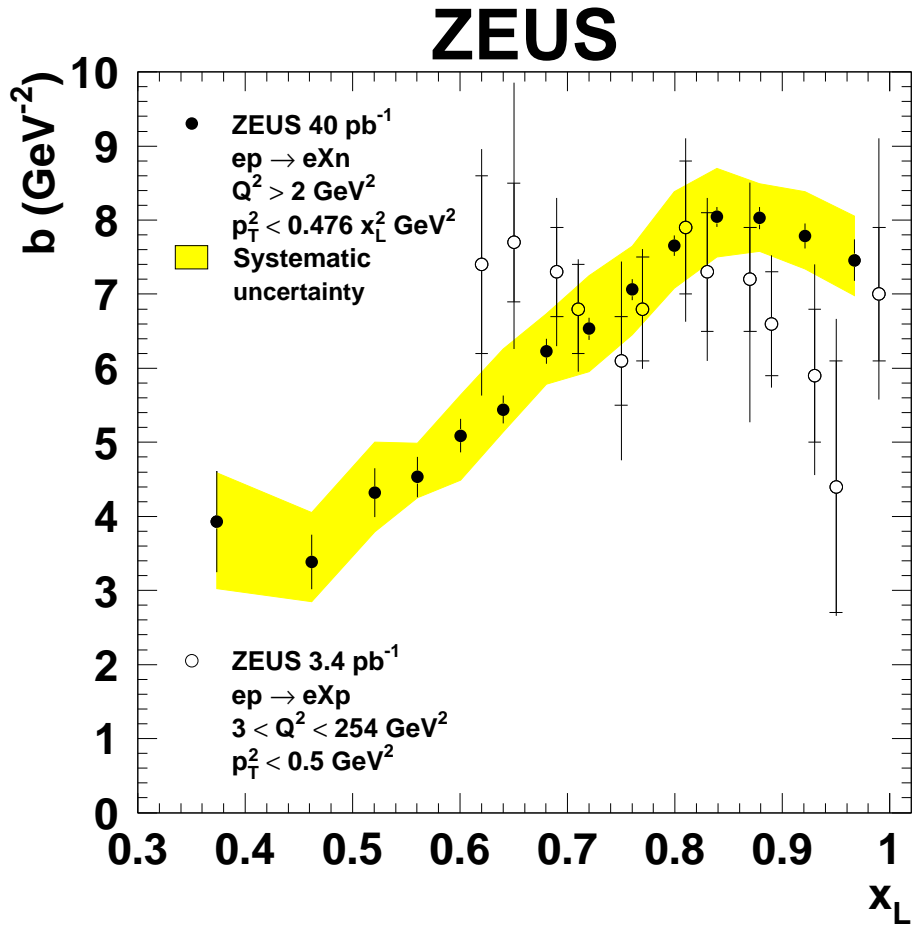,width=12cm}
\caption{
The exponential slopes of the $p_T^2$ distributions 
for leading neutrons and protons in DIS.
For the leading-neutron points, the error bars show the statistical
uncertainties;
the shaded band shows the systematic uncertainties.
For the leading-proton points, the inner error bars show
the statistical uncertainty only;
the full error bars show the statistical and systematic
uncertainties added in quadrature.
}
\label{fig-lpsb}
\end{figure}

The slopes of the $p_T^2$ distributions for leading protons
and neutrons in DIS are shown in \fig{lpsb}.
Note that the $p_T^2$ range for the proton measurement,
$p_T^2<0.5 \gev^2$,
is larger than for the neutron measurement.
The two samples have similar values of $b$ near $x_L \approx 0.8$, the region
where pion exchange is expected to
dominate for both processes~\cite{SNS}.

\section{Comparison to models}
\label{sec-modcomp}

In this section the data are compared to several models.
First the data are compared to various MC models
for the simulation of DIS events.
A comparison is then made to models incorporating only
pion exchange.
Next, more sophisticated models including the
effects of absorption of the neutron are considered.
Finally, a model incorporating enhanced absorption with pion and
additional secondary Regge exchanges is discussed.

\subsection{Monte Carlo models}

\begin{figure}[p]
\centering
\epsfig{file=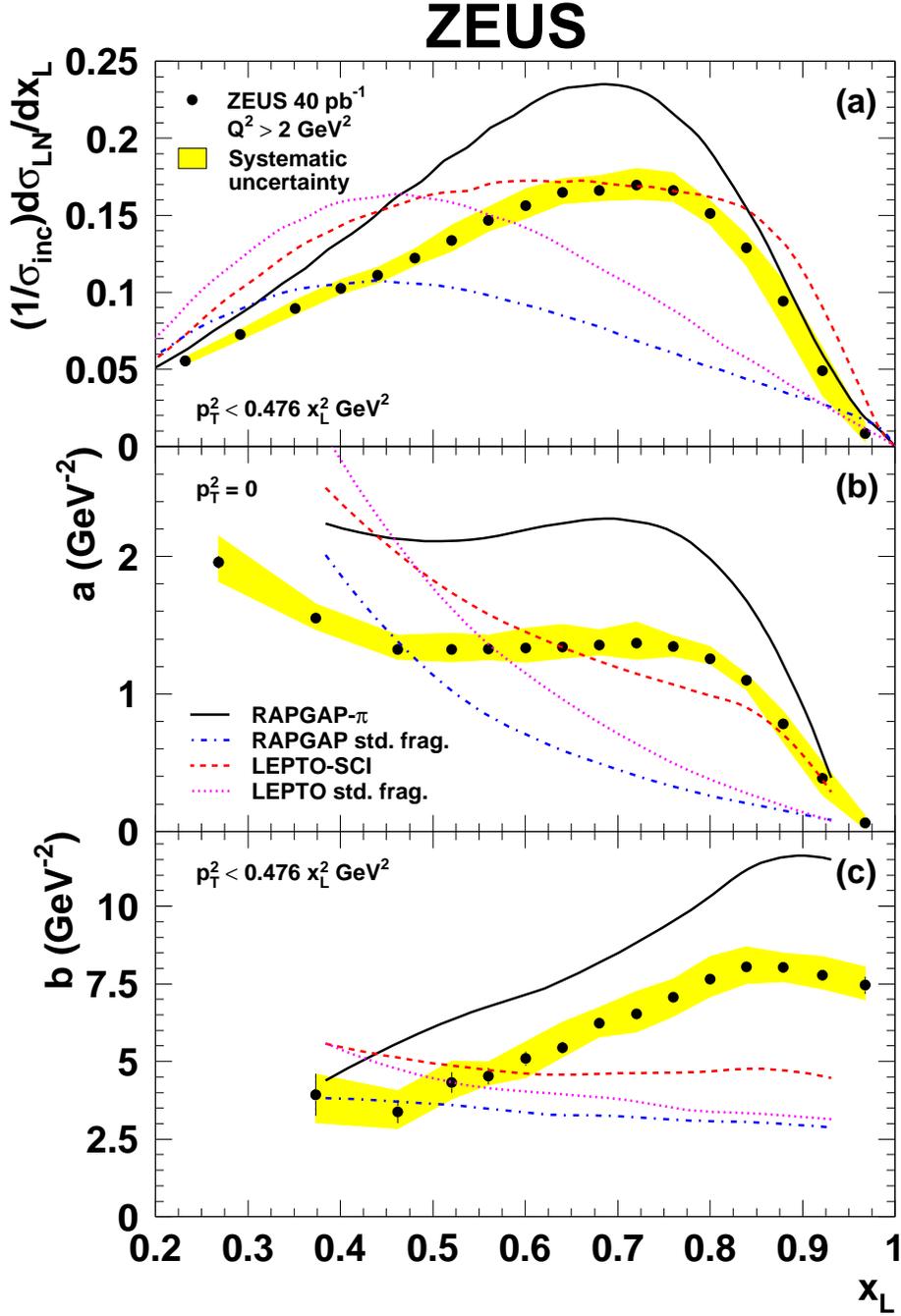,width=12cm}
\caption{
Comparison to Monte Carlo models of DIS:
(a) $x_L$ distributions,
(b) intercepts, and
(c) exponential slopes.
The error bars show the statistical uncertainties;
the shaded bands show the systematic uncertainties.
The bands do not include the overall normalization
uncertainty of $2.1\%$ for the $x_L$ distribution and intercepts.
The curves are from the DIS Monte Carlo models
{\sc Rapgap}~\pcite{cpc:86:147} and {\sc Lepto}~\pcite{cpc:101:108}.
The curves labeled {\sf RAPGAP std. frag.} and {\sf LEPTO std. frag.}
are the models incorporating only standard fragmentation.
The curve labeled \mbox{{\sf RAPGAP-}$\pi$} includes also
Pomeron and pion exchange;
the curve labeled {\sf LEPTO-SCI} is the model including
soft color interactions.
}
\label{fig-disraplep}
\end{figure}

Most MC models generate leading neutrons from
the fragmentation of the proton remnant~\cite{cpc:43:367}.
Some models also incorporate additional processes
to simulate diffraction and leading baryon production.
The leading neutron $x_L$ distribution, intercepts and slopes
in DIS are compared to two MC models in \fig{disraplep}.
The models presented here are
{\sc Rapgap}~\cite{cpc:86:147} and {\sc Lepto}~\cite{cpc:101:108}.
The proton PDF parameterization used was CTEQ5L~\cite{epj:c12:375}.
With only standard proton-remnant fragmentation,
the models are lower than the data in the
normalization of the $x_L$ distribution and intercepts, and are
peaked at lower $x_L$.
They do not show the observed $x_L$ dependence of the slopes.
Other models incorporating only standard fragmentation,
{\sc Ariadne}~\cite{cpc:71:15} and {\sc Cascade}~\cite{cpc:143:100} for DIS,
and {\sc Pythia}~\cite{cpc:82:74} and {\sc Phojet}~\cite{phojetref}
for photoproduction,
give a similarly poor description of the data.

{\sc Lepto} has the option to
implement soft color interactions (SCI)~\cite{pl:b366:371}
to produce rapidity gaps observed in diffractive events.
This model gives a qualitative description of the leading
proton $x_L$ distribution~\cite{np:b658:3}, including the diffractive peak,
although it predicts too few protons in the central $x_L$ region.
The predictions for leading neutrons,
with the probability of SCI set to 0.5,
 are shown in \fig{disraplep}.
{\sc Lepto} SCI comes close to the data in the shape and normalization
of the $x_L$ distribution.
The intercepts also exhibit a shoulder in the distribution
near $x_L \approx 0.8$ similar to that in the data.
It does not, however, give the observed strong $x_L$
dependence of the slopes. 

{\sc Rapgap} also includes Pomeron exchange to simulate diffractive
events, and pion exchange to simulate leading baryon production.
These processes are mixed with standard fragmentation
according to their respective cross sections.
The PDF parameterizations used here were 
the H1 LO fit 2~\cite{zfp:c76:613} for the Pomeron
and GRV-P LO fit~\cite{zfp:c53:651} for the pion.
The predictions for leading neutrons
from a mixture of these exchanges and standard fragmentation
are also shown in \fig{disraplep}.
The model well reproduces the shape of the $x_L$ distribution and
intercepts, although it predicts more neutrons than are observed.
The model also shows the strong $x_L$ dependence of the slopes in the data,
although the predicted values of the slopes are systematically
larger than the data.

\subsection{Pure pion exchange}

\begin{figure}[p]
\centering
\epsfig{file=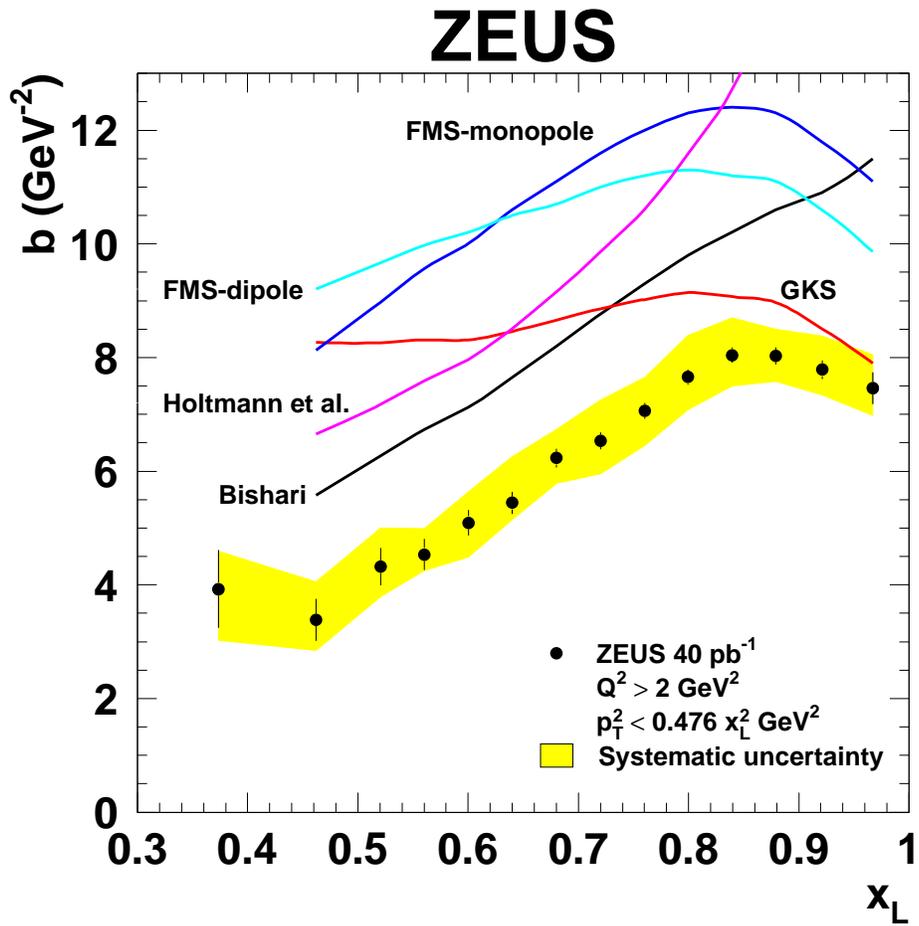,width=12cm}
\caption{
The measured exponential slopes $b$ compared to the predictions of 
models of one-pion exchange alone.
The error bars show the statistical
uncertainties;
the shaded band shows the systematic uncertainties.
The curves show predictions from the models discussed
in the text~\pcite{BISH,pl:b338:363,FMS,*GKS}.
}
\label{fig-bope}
\end{figure}

In the Regge factorization relation discussed in the introduction,
the leading neutron $x_L$ distribution is a product
of the pion flux, $f_{\pi/p}$,
and the $\gamma^* \pi $ cross section, $\sigma_{\gamma^* \pi}$.
However, if $\sigma_{\gamma^* \pi}$
is assumed to be independent of $t$,
the $p_T$ distribution of the produced neutrons is
determined only by the pion flux $f_{\pi/p}$.
The slopes can be compared to various parameterizations of the flux.
Although $f_{\pi/p}$ is not an exponential in $p_T^2$,
at fixed $x_L$ the models can be fit to the form
$\exp(-bp_T^2)$ using the same binning as the data,
and the resulting $b(x_L)$ values compared to the measurements.
All of the parameterizations in the 
literature~\cite{BISH,FMS,*GKS,pl:b338:363,KPP,*PSI,*MST,*SSS,SNS}
give values for the slopes larger than the data.
Most of them also have the wrong $x_L$ dependence of the slopes.
The models that most resemble the data
are shown in
\fig{bope}.
The simple model of Bishari~\cite{BISH}, with the form factor $F(x_L,t)=1$,
is closest in magnitude to the data.
Other models with more detailed parameterizations show
the turnover of the slopes near $x_L \approx 0.85$~\cite{FMS,*GKS}.
The model of
Holtmann et al.~\cite{pl:b338:363} is used for pion exchange in
the {\sc Rapgap}-$\pi$ MC.
The values for $b$ from {\sc Rapgap}-$\pi$ in \fig{disraplep} are lower
than this curve because {\sc Rapgap} also includes a contribution from
standard
fragmentation, which has flatter $p_T^2$ distributions
than pion exchange.
None of these models, based on pion exchange alone, describes the data.

\subsection{Pion exchange with neutron absorption}

\begin{figure}[p]
\centering
\epsfig{file=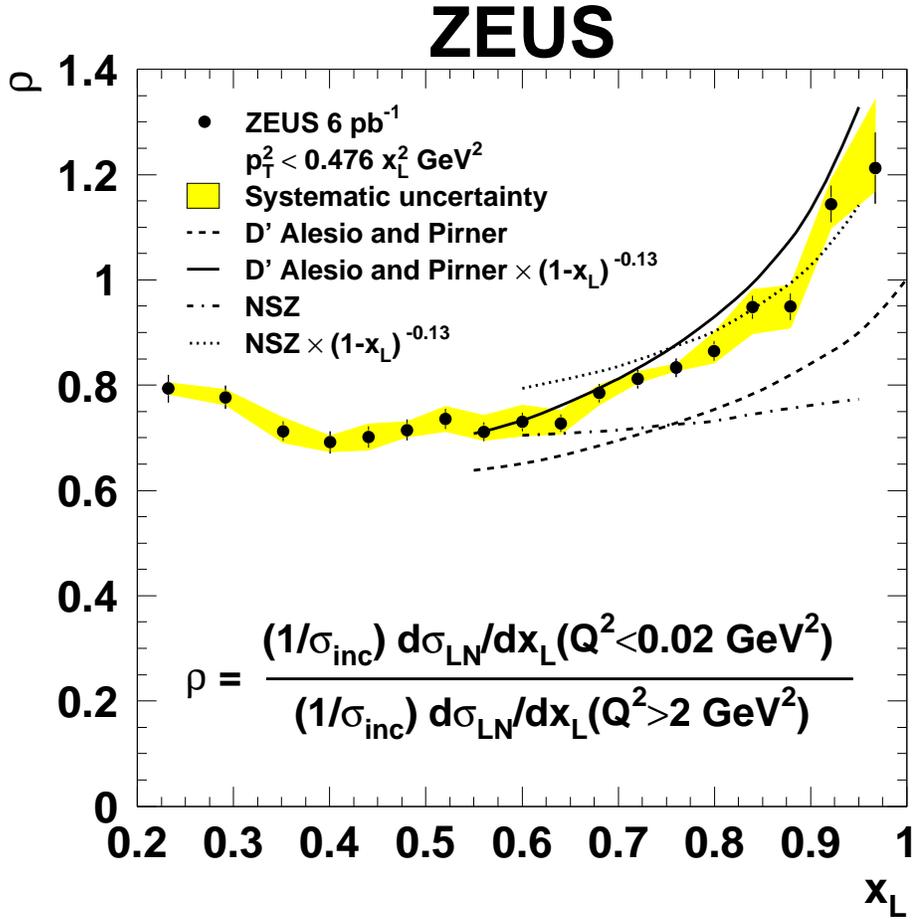,width=12cm}
\caption{
Ratio of photoproduction and DIS $x_L$ distributions.
The error bars show the statistical uncertainties;
the shaded band shows the systematic uncertainties.
The band does not include the overall normalization
uncertainty of $5.1\%$ on the ratio.
The dashed curve shows the neutron loss from a model
of rescattering~\pcite{\DAP}; the solid curve is this
model corrected for the different $W$ dependence of the
pion cross section in DIS and photoproduction.
The dot-dashed curve shows the neutron loss from
another absorption model\pcite{\NSZ};
the dotted curve is this
model corrected for the different $W$ dependence of the
pion cross section in DIS and photoproduction.
}
\label{fig-xlrhomods}
\end{figure}

As discussed in the introduction, in a geometrical picture,
neutron absorption may
occur for large photon sizes and small $n$-$\pi$ separations.
The former is inversely related to $Q^2$, and so is largest for
photoproduction; thus more absorption is expected in
photoproduction than in DIS.
The $n$-$\pi$ separation $r_{n \pi}$ is the Fourier conjugate of $p_T$,
and the distribution of $r_{n \pi}$ is given by
the Fourier transform of $f_{\pi/p}(p_T)$.
Parameterizations of the pion flux in general show that
the mean value of $r_{n \pi}$ increases with $x_L$,
so more absorption is expected at lower $x_L$ than at higher $x_L$.
The dashed curve in \fig{xlrhomods} is the expectation for the suppression
of leading neutrons in photoproduction relative to DIS from
a model of pion exchange with neutron absorption~\cite{\DAP}.
Although the curve lies below the data, it follows the same trend.
The $\gamma p$ interaction  has a power-law
dependence $\sigma \propto W^{2\lambda}$,
with different values of $\lambda$ for DIS and photoproduction.
Assuming that $\gamma \pi$ interactions have the same
dependence, and recalling that
$W_{\gamma \pi} = \sqrt{1-x_L} W_{\gamma p}$,
the ratio of photoproduction and
DIS cross sections is proportional to $(1-x_L)^{\Delta \lambda}$.
Previous ZEUS measurements of $\lambda$ in
photoproduction~\cite{np:b627:3}
and DIS~\cite{epj:c7:609,np:b713:3}
give $\Delta \lambda \approx -0.13$.
Applying this to the absorption suppression factor results in the
solid curve  in \fig{xlrhomods}.
Within the normalization uncertainty of $5.1\%$,
the data are well described by the absorption model with
this correction for different $W$ dependences.
Hence such a
geometric absorption model can account for the differences
between the $x_L$ distributions in DIS and photoproduction.

Also shown in \fig{xlrhomods} is
the NSZ model~\cite{\NSZ}
which employs the optical theorem together with multi-Pomeron
exchanges to describe all possible rescattering processes of the
leading hadron, resulting in absorptive effects.
With the correction for different $W$ dependences,
the prediction is close in magnitude to the data,
but does not have as steep an $x_L$ dependence.

\subsection{Enhanced neutron absorption and secondary exchanges}

\begin{figure}[p]
\centering
\epsfig{file=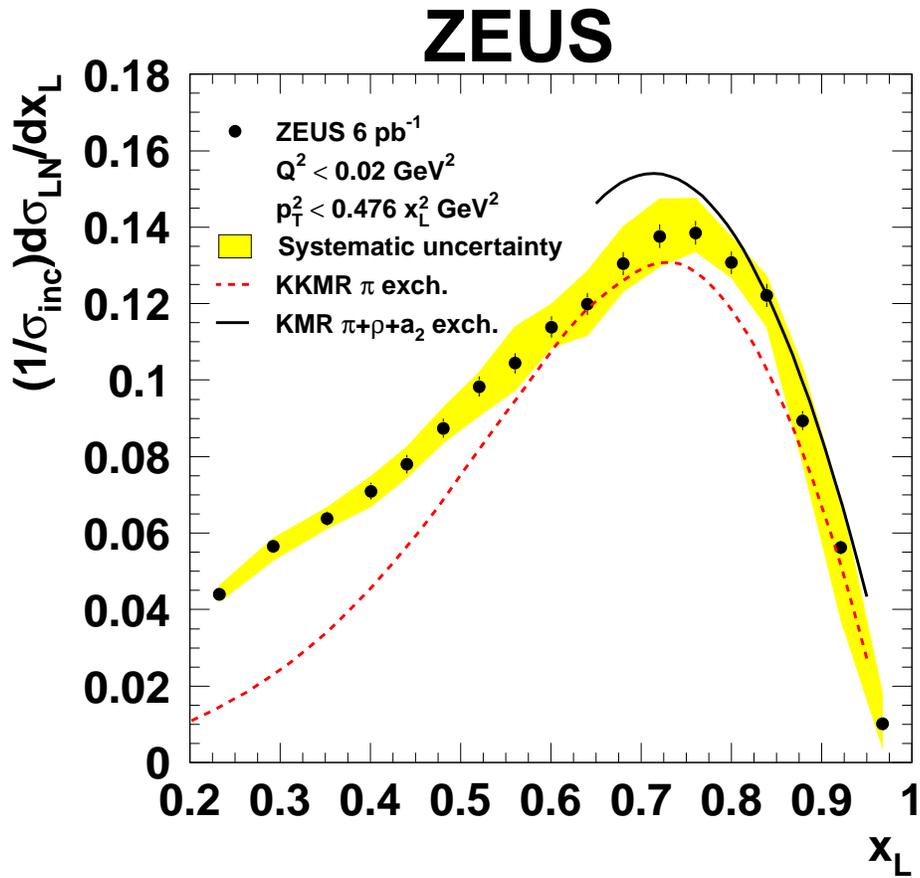,width=12cm}
\caption{
$x_L$ distribution for photoproduction. 
The error bars show the statistical uncertainties;
the shaded band shows the systematic uncertainties.
The band does not include the overall normalization uncertainty of
$5.5\%$.
The dashed curve shows the prediction of a model with enhanced neutron
absorption and migration for pion exchange
only~\pcite{epj:c47:385}.
The solid curve shows the same model including also
$\rho$ and $a_2$ exchanges~\pcite{Khoze:2006hwdo}.
}
\label{fig-xlphpKKMR}
\end{figure}

\begin{figure}[p]
\centering
\epsfig{file=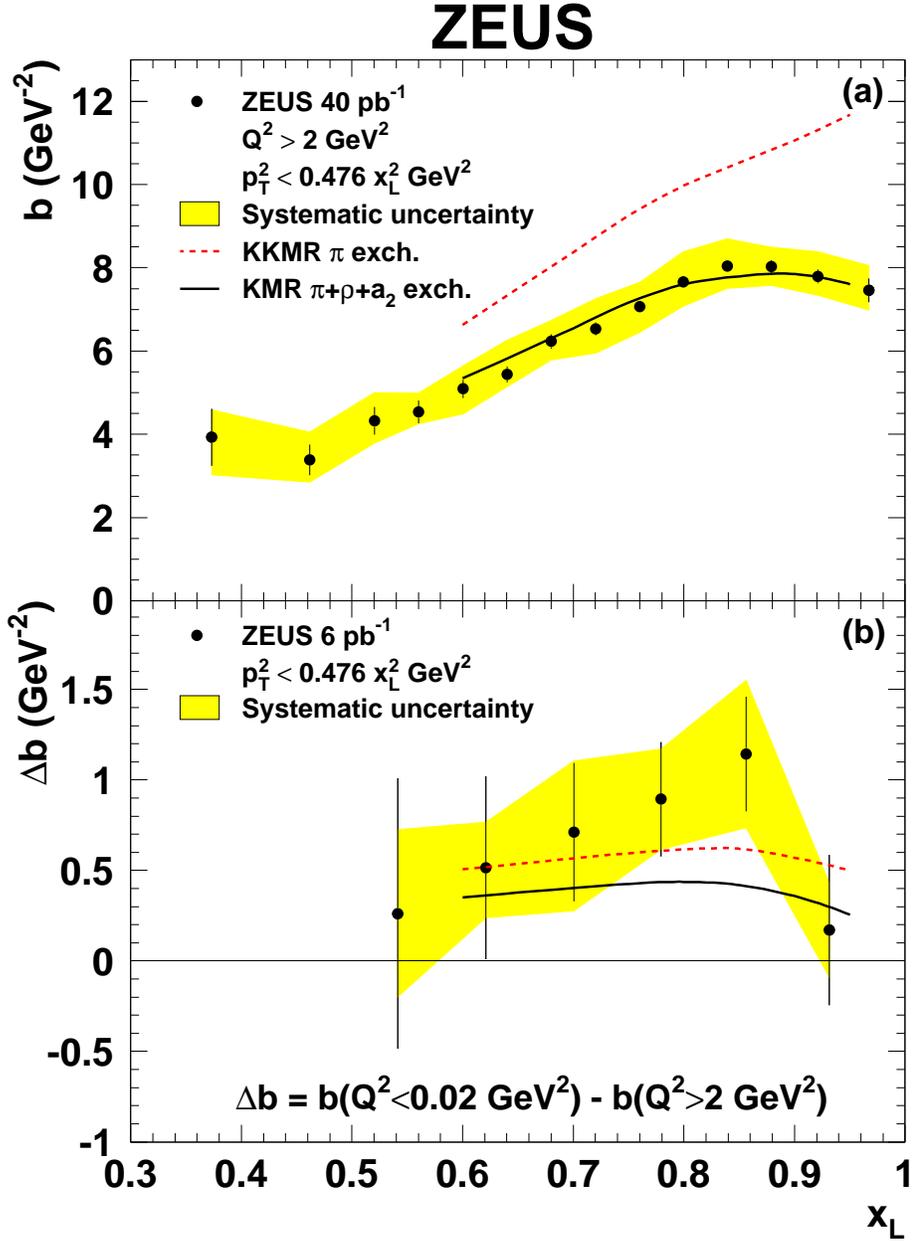,width=12cm}
\caption{
(a) Exponential slopes $b$ for DIS, and
(b) difference of exponential slopes $b$ for photoproduction and DIS.
In both plots
the error bars show the statistical uncertainties;
the shaded band shows the systematic uncertainties.
The dashed curves are from a model~\pcite{Khoze:2006hwdo}
based on pion exchange with enhanced neutron 
absorption and migration;
the solid curves include also
$\rho$ and $a_2$ exchanges.
}
\label{fig-bdbKMR}
\end{figure}

Recently a new calculation (KKMR)
of pion exchange with neutron
absorption based on multi-Pomeron exchanges
has become available~\cite{epj:c47:385}.
The pion exchange is based on the Bishari flux.
In addition to the rescatterings implemented in the earlier
model~\cite{\NSZ},
a small contribution from rescattering on intermediate partons
in the central rapidity region
is also included.
The model also accounts for the migration of neutrons
in $(x_L,p_T^2)$ after rescattering.
The prediction for the neutron $x_L$ distribution for
photoproduction, where rescattering is most important,
is shown by the dashed curve in \fig{xlphpKKMR}.
The model gives a fair description of both the
shape and normalization of the data.
The loss of neutrons through absorption is approximately $50\%$;
this is consistent with the deviation from the
$\sigma_{\gamma \pi}/\sigma_{\gamma p}=2/3$
prediction of the additive quark model
that was noted in the previous ZEUS measurement~\cite{np:b637:3}.
Within this model,
the present data can be used to constrain
the gap-survival probability,
one of the crucial inputs to calculations of diffractive
interactions at the LHC --- both hard, such as central
exclusive Higgs production, and soft, such as those giving rise
to the diffractive pile-up events~\cite{Khoze:2006reg}.
The prediction of this model for the slopes $b$ in DIS is shown
in \fig{bdbKMR}.
As for the pure pion-exchange calculations, the model predicts
larger values of $b$ than seen in the data.
This model does give a fair prediction for the magnitude
of the difference of the slopes in photoproduction and DIS, as
shown in \fig{bdbKMR}.

More recently this model was extended to include,
in addition to pion exchange,
the exchange of secondary $(\rho,a_2)$ Reggeons~\cite{Khoze:2006hwdo}.
This extended model gives a fair description of the shape and
normalization of the $x_L$ distribution in photoproduction,
as seen in \fig{xlphpKKMR}.
Since there are additional exchanges
the model gives a prediction for the $x_L$ distribution which
is higher than for pion-exchange alone.
As shown in \fig{bdbKMR}, the model
with secondary exchanges also gives a good prediction for the slopes.
Its description of the slope differences
is also close in magnitude to
the data, as seen in \fig{bdbKMR}.

\section{Summary}

The $x_L$ and $p_T^2$ distributions
of leading neutrons in photoproduction and DIS events at HERA
have been measured.
The $x_L$ distributions for the measured region
$\theta_n < 0.75 \mrad$
rise from the lowest $x_L$ due to the 
increasing $p_T$ phase space,
reach a maximum near $x_L=0.7$,
and fall to zero at the endpoint $x_L=1$.
The $p_T^2$ distributions are well described by an exponential
$d\sigma/dp_T^2 = a \exp(-bp_T^2)$.
The intercepts $a$ fall rapidly from the lowest $x_L$,
are roughly constant in the region $x_L=0.45$-$0.75$,
and fall to zero at the endpoint $x_L=1$.
The exponential slopes $b$ rise
linearly with $x_L$ in the range \mbox{$0.3 < x_L <0.85$} to a value of
$b \simeq 8 \gev^{-2}$, and then decrease slightly at higher $x_L$.

The neutron yield rises monotonically with $Q^2$ from the 
photoproduction region $Q^2<0.02\gev^2$ to the high-$Q^2$ DIS
region $Q^2>20\gev^2$.
The relative rise in yield
is greatest near $x_L=0.5$ and becomes
less significant at higher $x_L$.
The slopes of the $p_T^2$ distributions do not change significantly
within the DIS region $Q^2>2\gev^2$, but the slopes in
photoproduction exhibit a small
increase over
those in DIS for
$0.6<x_L<0.9$.

In the kinematic region $0.6<x_L<0.9$ and $p_T^2 < 0.04 \gev^2$
there are approximately twice as many leading protons as neutrons.
This indicates that leading proton production proceeds through
exchanges in addition to pure isovector (e.g. pion) exchange.
The slopes $b$ for leading protons 
agree with the
neutron slopes near $x_L \approx 0.8$ where pion exchange
is expected to dominate both processes.

Monte Carlo models commonly in use for the simulation
of DIS and photoproduction events which implement
standard fragmentation of the proton remnant
do not describe the leading-neutron data.
They predict fewer neutrons, concentrated at lower $x_L$.
They also predict smaller $p_T^2$ slopes and do not
have the strong $x_L$ dependence of the data.
The inclusion of soft color interactions gives a
reasonable description of the $x_L$ distributions, but
again fails to predict the $p_T^2$ slopes.
A mixture of processes including standard fragmentation, diffraction
and pion exchange gives a good description of the $x_L$ distributions
and the $x_L$ dependence of the slopes, although they
are larger than the data.

The measured $b(x_L)$ dependence in DIS has been compared with
various pion-exchange models.
All models give values larger than the data.
The simplest model is closest in magnitude to the data;
other models reproduce the measured shape of $b(x_L)$.

The $Q^2$  dependence of the neutron yield and $p_T^2$ slopes
is consistent with absorption models where neutrons
from pion exchange with smaller $n$-$\pi$ separations are lost
through rescattering on larger photons.
The photon size increases with decreasing $Q^2$, and
the mean  $n$-$\pi$ separation is smaller in the mid-$x_L$ range than
at higher $x_L$.
The result is a depletion of neutrons with decreasing $Q^2$,
with the depletion greater at mid-$x_L$ than at higher $x_L$,
as seen in the data.
The loss of neutrons with small $n$-$\pi$ separations, corresponding
to large $p_T^2$, also explains the larger $p_T^2$ slopes
measured in photoproduction than in DIS.

A model of neutron production through pion exchange, incorporating
enhanced neutron absorption and migration of the neutrons in
$(x_L,p_T^2)$ after rescattering, gives a fair description of
the shape and normalization of
the $x_L$ distributions in DIS and photoproduction, and of
the difference in the $p_T^2$ slopes $b$ between the two sets.
However, as with pure pion exchange, it predicts too high a
value for $b$.
Extending the model to include also $\rho$ and $a_2$ exchanges
still gives  a fair description of
the shape and normalization of
the $x_L$ distributions in DIS and photoproduction,
and also good descriptions of the $p_T^2$ slopes and of
the differences between the two sets.
 
\section*{Acknowledgements}
 
We are especially grateful to the
DESY Directorate whose encouragement and financial support made
possible the construction and installation of the FNC.
We are also happy to acknowledge the DESY accelerator group 
for allowing the installation of the FNC in close
proximity to the HERA machine components.
We also acknowledge the support of the DESY computing staff.

We thank V. Khoze, A. Martin and M. Ryskin for valuable discussions
and for providing the results of their calculations.

\newpage

\providecommand{\etal}{et al.\xspace}
\providecommand{\coll}{Coll.\xspace}
\catcode`\@=11
\def\@bibitem#1{%
\ifmc@bstsupport
  \mc@iftail{#1}%
    {;\newline\ignorespaces}%
    {\ifmc@first\else.\fi\orig@bibitem{#1}}
  \mc@firstfalse
\else
  \mc@iftail{#1}%
    {\ignorespaces}%
    {\orig@bibitem{#1}}%
\fi}%
\catcode`\@=12

\newpage


\begin{center}

\tablefirsthead{
\hline
$x_L$ range & $\langle x_L \rangle$ & $p_T^2(\gev^2)$ &
$1/\sigma_{\rm inc} \: d\sigma_{\rm LN}/dx_L dp_T^2 \: (\gev^{-2})$ \\
}

\tablehead{
\multicolumn{4}{l}{\tab{dndpdxldis} (cont.)} \\
\hline
$x_L$ range & $\langle x_L \rangle$ & $p_T^2(\gev^2)$ &
$1/\sigma_{\rm inc} \: d\sigma_{\rm LN}/dx_L dp_T^2 \: (\gev^{-2})$ \\
\hline
}

\tabletail{\hline}

\bottomcaption{
The normalized doubly differential distributions
$ (1/\sigma_{\rm inc}) d^2 \sigma_{\rm LN}/dx_L dp_T^2$
for the full DIS sample.
Only statistical uncertainties are shown.
}
\begin{supertabular}{|c|c|c|c|}
 \hline
  0.20-0.32 & 0.27 & 7.96 $\cdot 10^{-4}$ & 2.031 $\pm$ 0.064 \\
           &      & 2.51 $\cdot 10^{-3}$ & 1.881 $\pm$ 0.064 \\
           &      & 4.86 $\cdot 10^{-3}$ & 1.955 $\pm$ 0.070 \\
           &      & 7.95 $\cdot 10^{-3}$ & 2.002 $\pm$ 0.068 \\
           &      & 1.19 $\cdot 10^{-2}$ & 1.962 $\pm$ 0.063 \\
           &      & 1.65 $\cdot 10^{-2}$ & 2.001 $\pm$ 0.062 \\
 \hline
 0.32-0.42 & 0.37 & 1.99 $\cdot 10^{-3}$ & 1.541 $\pm$ 0.035 \\
           &      & 6.46 $\cdot 10^{-3}$ & 1.454 $\pm$ 0.038 \\
           &      & 1.24 $\cdot 10^{-2}$ & 1.445 $\pm$ 0.039 \\
           &      & 2.04 $\cdot 10^{-2}$ & 1.527 $\pm$ 0.037 \\
           &      & 3.03 $\cdot 10^{-2}$ & 1.440 $\pm$ 0.034 \\
           &      & 4.23 $\cdot 10^{-2}$ & 1.256 $\pm$ 0.031 \\
 \hline
 0.42-0.50 & 0.46 & 3.42 $\cdot 10^{-3}$ & 1.336 $\pm$ 0.026 \\
           &      & 1.11 $\cdot 10^{-2}$ & 1.211 $\pm$ 0.029 \\
           &      & 2.14 $\cdot 10^{-2}$ & 1.218 $\pm$ 0.029 \\
           &      & 3.51 $\cdot 10^{-2}$ & 1.217 $\pm$ 0.027 \\
           &      & 5.22 $\cdot 10^{-2}$ & 1.147 $\pm$ 0.025 \\
           &      & 7.28 $\cdot 10^{-2}$ & 1.005 $\pm$ 0.023 \\
 \hline
 0.50-0.54 & 0.52 & 4.84 $\cdot 10^{-3}$ & 1.274 $\pm$ 0.030 \\
           &      & 1.58 $\cdot 10^{-2}$ & 1.218 $\pm$ 0.035 \\
           &      & 3.03 $\cdot 10^{-2}$ & 1.200 $\pm$ 0.034 \\
           &      & 4.97 $\cdot 10^{-2}$ & 1.110 $\pm$ 0.031 \\
           &      & 7.40 $\cdot 10^{-2}$ & 0.946 $\pm$ 0.026 \\
           &      & 1.03 $\cdot 10^{-1}$ & 0.836 $\pm$ 0.026 \\
 \hline
 0.54-0.58 & 0.56 & 5.64 $\cdot 10^{-3}$ & 1.328 $\pm$ 0.028 \\
           &      & 1.84 $\cdot 10^{-2}$ & 1.144 $\pm$ 0.031 \\
           &      & 3.53 $\cdot 10^{-2}$ & 1.166 $\pm$ 0.031 \\
           &      & 5.80 $\cdot 10^{-2}$ & 1.016 $\pm$ 0.027 \\
           &      & 8.62 $\cdot 10^{-2}$ & 0.904 $\pm$ 0.024 \\
           &      & 1.20 $\cdot 10^{-1}$ & 0.767 $\pm$ 0.023 \\
  \end{supertabular}
  \label{tab-dndpdxldis}
\end{center}

\clearpage

\begin{center}

\tablefirsthead{
\multicolumn{4}{l}{\tab{dndpdxldis} (cont.)} \\
\hline
$x_L$ range & $\langle x_L \rangle$ & $p_T^2(\gev^2)$ &
$1/\sigma_{\rm inc} \: d\sigma_{\rm LN}/dx_L dp_T^2 \: (\gev^{-2})$ \\
}

\tablehead{
\multicolumn{4}{l}{\tab{dndpdxldis} (cont.)} \\
\hline
$x_L$ range & $\langle x_L \rangle$ & $p_T^2(\gev^2)$ &
$1/\sigma_{\rm inc} \: d\sigma_{\rm LN}/dx_L dp_T^2 \: (\gev^{-2})$ \\
}

\tabletail{\hline}

\begin{supertabular}{|c|c|c|c|}
 \hline
 0.58-0.62 & 0.60 & 6.50 $\cdot 10^{-3}$ & 1.286 $\pm$ 0.025 \\
           &      & 2.12 $\cdot 10^{-2}$ & 1.175 $\pm$ 0.029 \\
           &      & 4.07 $\cdot 10^{-2}$ & 1.112 $\pm$ 0.028 \\
           &      & 6.68 $\cdot 10^{-2}$ & 0.976 $\pm$ 0.024 \\
           &      & 9.94 $\cdot 10^{-2}$ & 0.795 $\pm$ 0.020 \\
           &      & 1.39 $\cdot 10^{-1}$ & 0.654 $\pm$ 0.019 \\
 \hline
 0.62-0.66 & 0.64 & 7.42 $\cdot 10^{-3}$ & 1.296 $\pm$ 0.024 \\
           &      & 2.42 $\cdot 10^{-2}$ & 1.157 $\pm$ 0.027 \\
           &      & 4.65 $\cdot 10^{-2}$ & 1.034 $\pm$ 0.025 \\
           &      & 7.63 $\cdot 10^{-2}$ & 0.926 $\pm$ 0.022 \\
           &      & 1.14 $\cdot 10^{-1}$ & 0.703 $\pm$ 0.018 \\
           &      & 1.58 $\cdot 10^{-1}$ & 0.571 $\pm$ 0.017 \\
 \hline
 0.66-0.70 & 0.68 & 8.39 $\cdot 10^{-3}$ & 1.297 $\pm$ 0.022 \\
           &      & 2.74 $\cdot 10^{-2}$ & 1.112 $\pm$ 0.025 \\
           &      & 5.27 $\cdot 10^{-2}$ & 0.990 $\pm$ 0.022 \\
           &      & 8.64 $\cdot 10^{-2}$ & 0.821 $\pm$ 0.019 \\
           &      & 1.29 $\cdot 10^{-1}$ & 0.581 $\pm$ 0.015 \\
           &      & 1.79 $\cdot 10^{-1}$ & 0.456 $\pm$ 0.014 \\
 \hline
 0.70-0.74 & 0.72 & 9.42 $\cdot 10^{-3}$ & 1.306 $\pm$ 0.021 \\
           &      & 3.08 $\cdot 10^{-2}$ & 1.126 $\pm$ 0.023 \\
           &      & 5.92 $\cdot 10^{-2}$ & 0.914 $\pm$ 0.020 \\
           &      & 9.71 $\cdot 10^{-2}$ & 0.707 $\pm$ 0.016 \\
           &      & 1.44 $\cdot 10^{-1}$ & 0.524 $\pm$ 0.013 \\
           &      & 2.01 $\cdot 10^{-1}$ & 0.388 $\pm$ 0.012 \\
 \hline
 0.74-0.78 & 0.76 & 1.05 $\cdot 10^{-2}$ & 1.280 $\pm$ 0.019 \\
           &      & 3.44 $\cdot 10^{-2}$ & 1.030 $\pm$ 0.021 \\
           &      & 6.61 $\cdot 10^{-2}$ & 0.837 $\pm$ 0.018 \\
           &      & 1.08 $\cdot 10^{-1}$ & 0.617 $\pm$ 0.014 \\
           &      & 1.61 $\cdot 10^{-1}$ & 0.403 $\pm$ 0.011 \\
           &      & 2.25 $\cdot 10^{-1}$ & 0.308 $\pm$ 0.010 \\
\hline
 0.78-0.82 & 0.80 & 1.16 $\cdot 10^{-2}$ & 1.180 $\pm$ 0.018 \\
           &      & 3.81 $\cdot 10^{-2}$ & 0.920 $\pm$ 0.019 \\
           &      & 7.33 $\cdot 10^{-2}$ & 0.714 $\pm$ 0.016 \\
           &      & 1.20 $\cdot 10^{-1}$ & 0.482 $\pm$ 0.012 \\
           &      & 1.79 $\cdot 10^{-1}$ & 0.304 $\pm$ 0.009 \\
           &      & 2.49 $\cdot 10^{-1}$ & 0.213 $\pm$ 0.008 \\
 \hline
 0.82-0.86 & 0.84 & 1.28 $\cdot 10^{-2}$ & 1.000 $\pm$ 0.016 \\
           &      & 4.21 $\cdot 10^{-2}$ & 0.783 $\pm$ 0.017 \\
           &      & 8.10 $\cdot 10^{-2}$ & 0.568 $\pm$ 0.014 \\
           &      & 1.33 $\cdot 10^{-1}$ & 0.374 $\pm$ 0.010 \\
           &      & 1.98 $\cdot 10^{-1}$ & 0.208 $\pm$ 0.007 \\
           &      & 2.75 $\cdot 10^{-1}$ & 0.139 $\pm$ 0.006 \\
 \hline
 0.86-0.90 & 0.88 & 1.41 $\cdot 10^{-2}$ & 0.719 $\pm$ 0.013 \\
           &      & 4.63 $\cdot 10^{-2}$ & 0.537 $\pm$ 0.013 \\
           &      & 8.90 $\cdot 10^{-2}$ & 0.363 $\pm$ 0.010 \\
           &      & 1.46 $\cdot 10^{-1}$ & 0.232 $\pm$ 0.007 \\
           &      & 2.17 $\cdot 10^{-1}$ & 0.135 $\pm$ 0.005 \\
           &      & 3.03 $\cdot 10^{-1}$ & 0.082 $\pm$ 0.004 \\
 \hline
 0.90-0.95 & 0.92 & 1.54 $\cdot 10^{-2}$ & 0.347 $\pm$ 0.007 \\
           &      & 5.07 $\cdot 10^{-2}$ & 0.267 $\pm$ 0.008 \\
           &      & 9.74 $\cdot 10^{-2}$ & 0.172 $\pm$ 0.006 \\
           &      & 1.60 $\cdot 10^{-1}$ & 0.110 $\pm$ 0.004 \\
           &      & 2.38 $\cdot 10^{-1}$ & 0.059 $\pm$ 0.003 \\
           &      & 3.31 $\cdot 10^{-1}$ & 0.034 $\pm$ 0.002 \\
 \hline
 0.95-1.00 & 0.97 & 1.72 $\cdot 10^{-2}$ & 0.054 $\pm$ 0.002 \\
           &      & 5.64 $\cdot 10^{-2}$ & 0.043 $\pm$ 0.002 \\
           &      & 1.08 $\cdot 10^{-1}$ & 0.032 $\pm$ 0.002 \\
           &      & 1.78 $\cdot 10^{-1}$ & 0.015 $\pm$ 0.001 \\
           &      & 2.65 $\cdot 10^{-1}$ & 0.009 $\pm$ 0.001 \\
           &      & 3.69 $\cdot 10^{-1}$ & 0.005 $\pm$ 0.001 \\
 \end{supertabular}
\end{center}

\clearpage

\begin{table}[p]
\begin{center}
\begin{tabular}{|c|c|c|c|}
\hline
$x_L$ range & $\langle x_L \rangle$ & $a \: (\gev^{-2})$ & $b \: (\gev^{-2})$  \\
\hline
 0.20-0.32 & 0.27 & $ 1.958 \pm 0.045^{+0.19}_{-0.14}$ & $ -0.94 \pm 2.40^{+2.19}_{-1.16}$ \\
 0.32-0.42 & 0.37 & $ 1.551 \pm 0.025^{+0.10}_{-0.08}$ & $ 3.93 \pm 0.68^{+0.66}_{-0.90}$ \\
 0.42-0.50 & 0.46 & $ 1.325 \pm 0.020^{+0.10}_{-0.07}$ & $ 3.38 \pm 0.37^{+0.67}_{-0.54}$ \\
 0.50-0.54 & 0.52 & $ 1.324 \pm 0.024^{+0.12}_{-0.09}$ & $ 4.32 \pm 0.33^{+0.68}_{-0.53}$ \\
 0.54-0.58 & 0.56 & $ 1.327 \pm 0.022^{+0.10}_{-0.08}$ & $ 4.53 \pm 0.27^{+0.45}_{-0.28}$ \\
 0.58-0.62 & 0.60 & $ 1.336 \pm 0.021^{+0.14}_{-0.10}$ & $ 5.09 \pm 0.22^{+0.55}_{-0.60}$ \\
 0.62-0.66 & 0.64 & $ 1.343 \pm 0.020^{+0.16}_{-0.08}$ & $ 5.44 \pm 0.19^{+0.82}_{-0.30}$ \\
 0.66-0.70 & 0.68 & $ 1.357 \pm 0.019^{+0.11}_{-0.07}$ & $ 6.23 \pm 0.17^{+0.50}_{-0.45}$ \\
 0.70-0.74 & 0.72 & $ 1.370 \pm 0.018^{+0.15}_{-0.12}$ & $ 6.53 \pm 0.15^{+0.72}_{-0.58}$ \\
 0.74-0.78 & 0.76 & $ 1.344 \pm 0.017^{+0.08}_{-0.07}$ & $ 7.06 \pm 0.14^{+0.59}_{-0.61}$ \\
 0.78-0.82 & 0.80 & $ 1.258 \pm 0.017^{+0.09}_{-0.04}$ & $ 7.66 \pm 0.14^{+0.73}_{-0.57}$ \\
 0.82-0.86 & 0.84 & $ 1.098 \pm 0.015^{+0.05}_{-0.07}$ & $ 8.04 \pm 0.14^{+0.65}_{-0.54}$ \\
 0.86-0.90 & 0.88 & $ 0.782 \pm 0.013^{+0.09}_{-0.13}$ & $ 8.03 \pm 0.15^{+0.46}_{-0.45}$ \\
 0.90-0.95 & 0.92 & $ 0.387 \pm 0.007^{+0.12}_{-0.12}$ & $ 7.79 \pm 0.17^{+0.60}_{-0.45}$ \\
 0.95-1.00 & 0.97 & $ 0.063 \pm 0.002^{+0.05}_{-0.04}$ & $ 7.46 \pm 0.28^{+0.59}_{-0.47}$ \\
\hline
\end{tabular}
\caption{
The intercepts $a$ and slopes $b$ from the exponential
parameterization of the differential cross defined
in \Sect{res} for the full DIS sample.
Statistical uncertainties are listed first, followed by
systematic uncertainties, not including an overall
normalization uncertainty of 2.1\% on the intercepts.
The systematic uncertainties are largely correlated
between all points.
}
  \label{tab-abdis}
\end{center}
\end{table}                                                                     

\clearpage

\begin{table}[p]
\begin{center}
\begin{tabular}{|c|c|c|c|c|}
\hline
sample & $Q^2$ range $(\gev^2)$& $\langle Q^2 \rangle$ $(\gev^2)$ 
            & $W$ range $(\gev)$& $\langle W \rangle$ $(\gev)$\\ 
\hline
\hline
$\gamma p$ & $Q^2<0.02$ & $4 \times 10^{-4}$ & $150<W<270$ & $215$ \\
\hline
full DIS & $Q^2>2$ & $13$ & $W<250$ & $95$ \\
\hline
low-$Q^2$ DIS & $2<Q_{DA}^2<5$ & $2.7$ & $50<W<250$ & $140$ \\
\hline
mid-$Q^2$ DIS & $5<Q_{DA}^2<20$ & $8.9$ & $50<W<250$ & $132$ \\
\hline
high-$Q^2$ DIS & $20<Q_{DA}^2<120$ & $40$ & $50<W<250$ & $131$ \\
\hline
\end{tabular}
\caption{
Kinematic regions for each of the data samples.
The $\gamma p$ ranges were estimated from a simulation
of photoproduction including the positron tagger
used to trigger these data.
The $Q^2$ ranges for the low-, mid-, and high-$Q^2$ DIS samples
are the limits on the double angle variable $Q_{DA}^2$.
All other ranges and means for DIS were estimated using
the DJANGOH 1.1~\pcite{proc:hera:1991:1419} generator,
where the CTEQ4D~\pcite{pr:d55:1280} parton-density
parameterizations were used.
}
  \label{tab-kinranges}
\end{center}
\end{table}

\begin{table}[p]
\begin{center}
\begin{tabular}{|c|c|c|c|}
\hline
sample & $r_{\rm LN}$ \\
\hline
\hline
    $\gamma p$ & 0.0700  $\pm 0.0004$  $^{+0.0040}_{-0.0040}$ \\
\hline
 full DIS & 0.0885  $\pm 0.0002$  $^{+0.0029}_{-0.0022}$ \\
\hline
 low-$Q^2$ DIS & 0.0837  $\pm 0.0003$  $^{+0.0020}_{-0.0020}$ \\
\hline
 mid-$Q^2$ DIS & 0.0843  $\pm 0.0003$  $^{+0.0020}_{-0.0019}$ \\
\hline
high-$Q^2$ DIS & 0.0913  $\pm 0.0006$  $^{+0.0021}_{-0.0021}$ \\
\hline
\end{tabular}
\caption{
$r_{\rm LN}$ in the region $x_L>0.2, p_T^2<0.476 x_L^2 \gev^2$,
for the photoproduction and 
DIS samples.
Statistical uncertainties are listed first, followed by
systematic uncertainties, not including an overall
normalization uncertainty of 2.1\%.
}
  \label{tab-intxl}
\end{center}
\end{table}

\clearpage

\begin{center}

\tablefirsthead{
\hline
 & & &
\multicolumn{4}{c|}{$ 1/\sigma_{\rm inc} \: d\sigma_{\rm LN}/dx_L dp_T^2 \:
  (\gev^{-2})$} \\ \cline{4-7}
$x_L$ range & $\langle x_L \rangle$ & $p_T^2(\gev^2)$ & $\gamma p$ &
low-$Q^2$ DIS & mid-$Q^2$ DIS & high-$Q^2$ DIS \\
}

\tablehead{
\multicolumn{7}{l}{\tab{dndpdxlqq} (cont.)} \\
\hline
 & & &
\multicolumn{4}{c|}{$ 1/\sigma_{\rm inc} \: d\sigma_{\rm LN}/dx_L dp_T^2 \:
  (\gev^{-2})$} \\ \cline{4-7}
$x_L$ range & $\langle x_L \rangle$ & $p_T^2(\gev^2)$ & $\gamma p$ &
low-$Q^2$ DIS & mid-$Q^2$ DIS & high-$Q^2$ DIS \\
}

\tabletail{\hline}

\bottomcaption{ 
The normalized doubly differential distributions
$ (1/\sigma_{\rm inc}) d^2 \sigma_{\rm LN}/dx_L dp_T^2$
for the photoproduction and three DIS samples.
Only statistical uncertainties are shown.
}
{\footnotesize
\begin{supertabular}{|c|c|c|c|c|c|c|}
 \hline

 0.20-0.50 & 0.38 & 8.40 $\cdot 10^{-4}$ & 1.501 $\pm$ 0.108 & 1.491 $\pm$ 0.058 & 1.721 $\pm$ 0.063 & 1.848 $\pm$ 0.115 \\
           &      & 2.43 $\cdot 10^{-3}$ & 1.289 $\pm$ 0.090 & 1.342 $\pm$ 0.052 & 1.500 $\pm$ 0.056 & 1.749 $\pm$ 0.106 \\
           &      & 4.86 $\cdot 10^{-3}$ & 1.127 $\pm$ 0.076 & 1.370 $\pm$ 0.052 & 1.451 $\pm$ 0.053 & 1.757 $\pm$ 0.104 \\
           &      & 7.94 $\cdot 10^{-3}$ & 1.232 $\pm$ 0.084 & 1.417 $\pm$ 0.054 & 1.528 $\pm$ 0.057 & 1.809 $\pm$ 0.110 \\
           &      & 1.19 $\cdot 10^{-2}$ & 1.064 $\pm$ 0.075 & 1.359 $\pm$ 0.053 & 1.451 $\pm$ 0.055 & 1.803 $\pm$ 0.109 \\
           &      & 1.65 $\cdot 10^{-2}$ & 1.021 $\pm$ 0.068 & 1.374 $\pm$ 0.052 & 1.589 $\pm$ 0.056 & 1.805 $\pm$ 0.106 \\
 \hline
 0.50-0.58 & 0.54 & 4.84 $\cdot 10^{-3}$ & 0.988 $\pm$ 0.056 & 1.196 $\pm$ 0.036 & 1.247 $\pm$ 0.037 & 1.452 $\pm$ 0.071 \\
           &      & 1.58 $\cdot 10^{-2}$ & 0.791 $\pm$ 0.055 & 1.077 $\pm$ 0.042 & 1.096 $\pm$ 0.042 & 1.220 $\pm$ 0.079 \\
           &      & 3.03 $\cdot 10^{-2}$ & 0.812 $\pm$ 0.058 & 1.027 $\pm$ 0.040 & 1.042 $\pm$ 0.041 & 1.191 $\pm$ 0.078 \\
           &      & 4.97 $\cdot 10^{-2}$ & 0.784 $\pm$ 0.052 & 1.005 $\pm$ 0.037 & 1.009 $\pm$ 0.037 & 1.187 $\pm$ 0.072 \\
           &      & 7.39 $\cdot 10^{-2}$ & 0.654 $\pm$ 0.043 & 0.880 $\pm$ 0.033 & 0.857 $\pm$ 0.032 & 0.934 $\pm$ 0.060 \\
           &      & 1.03 $\cdot 10^{-1}$ & 0.591 $\pm$ 0.041 & 0.768 $\pm$ 0.031 & 0.760 $\pm$ 0.031 & 0.782 $\pm$ 0.056 \\
 \hline
 0.58-0.66 & 0.62 & 6.50 $\cdot 10^{-3}$ & 0.991 $\pm$ 0.047 & 1.192 $\pm$ 0.031 & 1.261 $\pm$ 0.032 & 1.221 $\pm$ 0.055 \\
           &      & 2.12 $\cdot 10^{-2}$ & 0.849 $\pm$ 0.051 & 1.062 $\pm$ 0.035 & 1.122 $\pm$ 0.037 & 1.131 $\pm$ 0.065 \\
           &      & 4.08 $\cdot 10^{-2}$ & 0.892 $\pm$ 0.055 & 1.025 $\pm$ 0.034 & 1.068 $\pm$ 0.035 & 1.151 $\pm$ 0.065 \\
           &      & 6.68 $\cdot 10^{-2}$ & 0.753 $\pm$ 0.045 & 0.906 $\pm$ 0.030 & 0.889 $\pm$ 0.030 & 1.090 $\pm$ 0.059 \\
           &      & 9.94 $\cdot 10^{-2}$ & 0.563 $\pm$ 0.035 & 0.714 $\pm$ 0.025 & 0.748 $\pm$ 0.025 & 0.837 $\pm$ 0.048 \\
           &      & 1.39 $\cdot 10^{-1}$ & 0.451 $\pm$ 0.031 & 0.575 $\pm$ 0.023 & 0.605 $\pm$ 0.023 & 0.626 $\pm$ 0.042 \\
 \hline
 0.66-0.74 & 0.70 & 8.38 $\cdot 10^{-3}$ & 1.083 $\pm$ 0.044 & 1.303 $\pm$ 0.028 & 1.263 $\pm$ 0.028 & 1.309 $\pm$ 0.050 \\
           &      & 2.74 $\cdot 10^{-2}$ & 0.882 $\pm$ 0.047 & 1.116 $\pm$ 0.032 & 1.093 $\pm$ 0.031 & 1.104 $\pm$ 0.056 \\
           &      & 5.27 $\cdot 10^{-2}$ & 0.815 $\pm$ 0.045 & 0.935 $\pm$ 0.028 & 0.961 $\pm$ 0.029 & 1.035 $\pm$ 0.054 \\
           &      & 8.64 $\cdot 10^{-2}$ & 0.688 $\pm$ 0.039 & 0.751 $\pm$ 0.024 & 0.739 $\pm$ 0.023 & 0.745 $\pm$ 0.042 \\
           &      & 1.28 $\cdot 10^{-1}$ & 0.451 $\pm$ 0.027 & 0.530 $\pm$ 0.018 & 0.560 $\pm$ 0.019 & 0.532 $\pm$ 0.033 \\
           &      & 1.79 $\cdot 10^{-1}$ & 0.305 $\pm$ 0.021 & 0.426 $\pm$ 0.017 & 0.428 $\pm$ 0.017 & 0.448 $\pm$ 0.031 \\

 \end{supertabular}}
  \label{tab-dndpdxlqq}
\end{center}

\clearpage
\begin{center}

\tablefirsthead{
\multicolumn{7}{l}{\tab{dndpdxlqq} (cont.)} \\
\hline
 & & &
\multicolumn{4}{c|}{$ 1/\sigma_{\rm inc} \: d\sigma_{\rm LN}/dx_L dp_T^2 \:
  (\gev^{-2})$} \\ \cline{4-7}
$x_L$ range & $\langle x_L \rangle$ & $p_T^2(\gev^2)$ & $\gamma p$ &
low-$Q^2$ DIS & mid-$Q^2$ DIS & high-$Q^2$ DIS \\
}

\tablehead{
\multicolumn{7}{l}{\tab{dndpdxlqq} (cont.)} \\
\hline
 & & &
\multicolumn{4}{c|}{$ 1/\sigma_{\rm inc} \: d\sigma_{\rm LN}/dx_L dp_T^2 \:
  (\gev^{-2})$} \\ \cline{4-7}
$x_L$ range & $\langle x_L \rangle$ & $p_T^2(\gev^2)$ & $\gamma p$ &
low-$Q^2$ DIS & mid-$Q^2$ DIS & high-$Q^2$ DIS \\
}

\tabletail{\hline}

{\footnotesize
\begin{supertabular}{|c|c|c|c|c|c|c|}
 \hline
 0.74-0.82 & 0.78 & 1.05 $\cdot 10^{-2}$ & 1.111 $\pm$ 0.041 & 1.250 $\pm$ 0.025 & 1.181 $\pm$ 0.024 & 1.251 $\pm$ 0.044 \\
           &      & 3.44 $\cdot 10^{-2}$ & 0.884 $\pm$ 0.043 & 1.027 $\pm$ 0.027 & 0.986 $\pm$ 0.027 & 0.915 $\pm$ 0.046 \\
           &      & 6.61 $\cdot 10^{-2}$ & 0.769 $\pm$ 0.042 & 0.824 $\pm$ 0.024 & 0.779 $\pm$ 0.023 & 0.846 $\pm$ 0.043 \\
           &      & 1.08 $\cdot 10^{-1}$ & 0.494 $\pm$ 0.028 & 0.587 $\pm$ 0.018 & 0.541 $\pm$ 0.018 & 0.576 $\pm$ 0.033 \\
           &      & 1.61 $\cdot 10^{-1}$ & 0.285 $\pm$ 0.018 & 0.361 $\pm$ 0.013 & 0.362 $\pm$ 0.013 & 0.399 $\pm$ 0.025 \\
           &      & 2.25 $\cdot 10^{-1}$ & 0.209 $\pm$ 0.016 & 0.267 $\pm$ 0.012 & 0.243 $\pm$ 0.011 & 0.289 $\pm$ 0.022 \\
 \hline
 0.82-0.90 & 0.86 & 1.28 $\cdot 10^{-2}$ & 0.876 $\pm$ 0.034 & 0.899 $\pm$ 0.019 & 0.894 $\pm$ 0.019 & 0.838 $\pm$ 0.032 \\
           &      & 4.21 $\cdot 10^{-2}$ & 0.688 $\pm$ 0.036 & 0.732 $\pm$ 0.021 & 0.671 $\pm$ 0.020 & 0.647 $\pm$ 0.035 \\
           &      & 8.09 $\cdot 10^{-2}$ & 0.538 $\pm$ 0.033 & 0.528 $\pm$ 0.017 & 0.477 $\pm$ 0.016 & 0.477 $\pm$ 0.029 \\
           &      & 1.33 $\cdot 10^{-1}$ & 0.293 $\pm$ 0.020 & 0.335 $\pm$ 0.013 & 0.329 $\pm$ 0.012 & 0.268 $\pm$ 0.020 \\
           &      & 1.97 $\cdot 10^{-1}$ & 0.134 $\pm$ 0.011 & 0.195 $\pm$ 0.009 & 0.170 $\pm$ 0.008 & 0.174 $\pm$ 0.015 \\
           &      & 2.75 $\cdot 10^{-1}$ & 0.098 $\pm$ 0.010 & 0.116 $\pm$ 0.007 & 0.113 $\pm$ 0.007 & 0.116 $\pm$ 0.012 \\
 \hline
 0.90-1.00 & 0.93 & 1.54 $\cdot 10^{-2}$ & 0.242 $\pm$ 0.015 & 0.236 $\pm$ 0.008 & 0.228 $\pm$ 0.007 & 0.201 $\pm$ 0.012 \\
           &      & 5.06 $\cdot 10^{-2}$ & 0.166 $\pm$ 0.014 & 0.180 $\pm$ 0.008 & 0.156 $\pm$ 0.007 & 0.142 $\pm$ 0.013 \\
           &      & 9.74 $\cdot 10^{-2}$ & 0.130 $\pm$ 0.012 & 0.117 $\pm$ 0.006 & 0.115 $\pm$ 0.006 & 0.093 $\pm$ 0.010 \\
           &      & 1.60 $\cdot 10^{-1}$ & 0.072 $\pm$ 0.008 & 0.077 $\pm$ 0.005 & 0.065 $\pm$ 0.004 & 0.057 $\pm$ 0.007 \\
           &      & 2.38 $\cdot 10^{-1}$ & 0.039 $\pm$ 0.005 & 0.041 $\pm$ 0.003 & 0.035 $\pm$ 0.003 & 0.034 $\pm$ 0.005 \\
           &      & 3.31 $\cdot 10^{-1}$ & 0.022 $\pm$ 0.004 & 0.019 $\pm$ 0.002 & 0.023 $\pm$ 0.002 & 0.019 $\pm$ 0.004 \\

 \end{supertabular}}
\end{center}

\clearpage
 
\begin{table}[p]
\begin{center}
\begin{tabular}{|c|c|c|c|c|}
\hline
 sample & $x_L$ range & $\langle x_L \rangle$ & $a \: (\gev^{-2})$ & $b \: (\gev^{-2})$  \\
\hline
$\gamma p$  & 0.20-0.50 & 0.38 & $ 1.386 \pm 0.068^{+0.10}_{-0.10}$ & $ 20.32 \pm 5.27^{+1.68}_{-2.44}$ \\
            & 0.50-0.58 & 0.54 & $ 0.949 \pm 0.041^{+0.07}_{-0.10}$ & $  4.72 \pm 0.78^{+0.73}_{-0.68}$ \\
            & 0.58-0.66 & 0.62 & $ 1.037 \pm 0.039^{+0.13}_{-0.08}$ & $  5.82 \pm 0.53^{+0.76}_{-0.65}$ \\
            & 0.66-0.74 & 0.70 & $ 1.149 \pm 0.038^{+0.11}_{-0.11}$ & $  7.13 \pm 0.40^{+0.70}_{-0.78}$ \\
            & 0.74-0.82 & 0.78 & $ 1.217 \pm 0.038^{+0.10}_{-0.05}$ & $  8.32 \pm 0.33^{+0.75}_{-0.55}$ \\
            & 0.82-0.90 & 0.86 & $ 1.009 \pm 0.035^{+0.04}_{-0.09}$ & $  9.23 \pm 0.33^{+0.50}_{-0.64}$ \\
            & 0.90-1.00 & 0.93 & $ 0.266 \pm 0.014^{+0.09}_{-0.10}$ & $  7.91 \pm 0.44^{+0.64}_{-0.41}$ \\
 \hline
low-$Q^2$   & 0.20-0.50 & 0.38 & $ 1.412 \pm 0.038^{+0.10}_{-0.09}$ & $  2.21 \pm 2.92^{+2.17}_{-2.18}$ \\
DIS         & 0.50-0.58 & 0.54 & $ 1.196 \pm 0.029^{+0.07}_{-0.07}$ & $  4.20 \pm 0.44^{+0.51}_{-0.40}$ \\
            & 0.58-0.66 & 0.62 & $ 1.238 \pm 0.025^{+0.13}_{-0.09}$ & $  5.40 \pm 0.30^{+0.95}_{-0.70}$ \\
            & 0.66-0.74 & 0.70 & $ 1.359 \pm 0.025^{+0.09}_{-0.13}$ & $  6.88 \pm 0.23^{+0.42}_{-0.73}$ \\
            & 0.74-0.82 & 0.78 & $ 1.346 \pm 0.023^{+0.05}_{-0.06}$ & $  7.65 \pm 0.19^{+0.49}_{-0.74}$ \\
            & 0.82-0.90 & 0.86 & $ 1.005 \pm 0.019^{+0.06}_{-0.10}$ & $  8.10 \pm 0.18^{+0.47}_{-0.53}$ \\
            & 0.90-1.00 & 0.93 & $ 0.266 \pm 0.008^{+0.10}_{-0.08}$ & $  7.94 \pm 0.25^{+0.41}_{-0.55}$ \\
 \hline
mid-$Q^2$   & 0.20-0.50 & 0.38 & $ 1.552 \pm 0.041^{+0.12}_{-0.08}$ & $  1.80 \pm 2.87^{+2.87}_{-2.18}$ \\
DIS         & 0.50-0.58 & 0.54 & $ 1.238 \pm 0.029^{+0.10}_{-0.09}$ & $  4.80 \pm 0.44^{+0.51}_{-0.47}$ \\
            & 0.58-0.66 & 0.62 & $ 1.298 \pm 0.026^{+0.10}_{-0.10}$ & $  5.53 \pm 0.30^{+0.54}_{-0.42}$ \\
            & 0.66-0.74 & 0.70 & $ 1.324 \pm 0.024^{+0.09}_{-0.10}$ & $  6.51 \pm 0.22^{+0.61}_{-0.49}$ \\
            & 0.74-0.82 & 0.78 & $ 1.276 \pm 0.022^{+0.07}_{-0.07}$ & $  7.65 \pm 0.19^{+0.57}_{-0.58}$ \\
            & 0.82-0.90 & 0.86 & $ 0.975 \pm 0.019^{+0.05}_{-0.08}$ & $  8.40 \pm 0.19^{+0.54}_{-0.54}$ \\
            & 0.90-1.00 & 0.93 & $ 0.248 \pm 0.008^{+0.10}_{-0.09}$ & $  7.99 \pm 0.28^{+0.45}_{-0.54}$ \\
 \hline
 high-$Q^2$  & 0.20-0.50 & 0.38 & $ 1.786 \pm 0.076^{+0.13}_{-0.15}$ & $ -0.53 \pm 4.56^{+4.71}_{-2.92}$ \\
 DIS        & 0.50-0.58 & 0.54 & $ 1.449 \pm 0.056^{+0.05}_{-0.10}$ & $  5.79 \pm 0.74^{+0.61}_{-0.78}$ \\
            & 0.58-0.66 & 0.62 & $ 1.309 \pm 0.045^{+0.15}_{-0.11}$ & $  4.66 \pm 0.48^{+0.92}_{-0.69}$ \\
            & 0.66-0.74 & 0.70 & $ 1.374 \pm 0.044^{+0.11}_{-0.05}$ & $  6.76 \pm 0.40^{+0.55}_{-0.63}$ \\
            & 0.74-0.82 & 0.78 & $ 1.283 \pm 0.039^{+0.03}_{-0.07}$ & $  7.02 \pm 0.33^{+0.78}_{-0.58}$ \\
            & 0.82-0.90 & 0.86 & $ 0.915 \pm 0.032^{+0.05}_{-0.09}$ & $  8.29 \pm 0.36^{+0.61}_{-0.71}$ \\
            & 0.90-1.00 & 0.93 & $ 0.217 \pm 0.013^{+0.10}_{-0.08}$ & $  7.93 \pm 0.53^{+0.38}_{-0.53}$ \\
 \hline
\end{tabular}
\caption{ 
The intercepts $a$ and slopes $b$ from the exponential
parameterization of the differential cross defined
in \Sect{res} for the photoproduction and three DIS samples.
Statistical uncertainties are listed first, followed by
systematic uncertainties, not including an overall
normalization uncertainty of 2.1\% on the intercepts in DIS,
nor an additional uncorrelated uncertainty of
5.1\% on the photoproduction intercepts.
The systematic uncertainties are largely correlated
between all points.
}
  \label{tab-abqq}
\end{center}
\end{table}

\pagestyle{plain}

\clearpage


\clearpage

\clearpage

\clearpage

\clearpage

\clearpage

\clearpage

\begin{figure}[p]
\centering
\epsfig{file=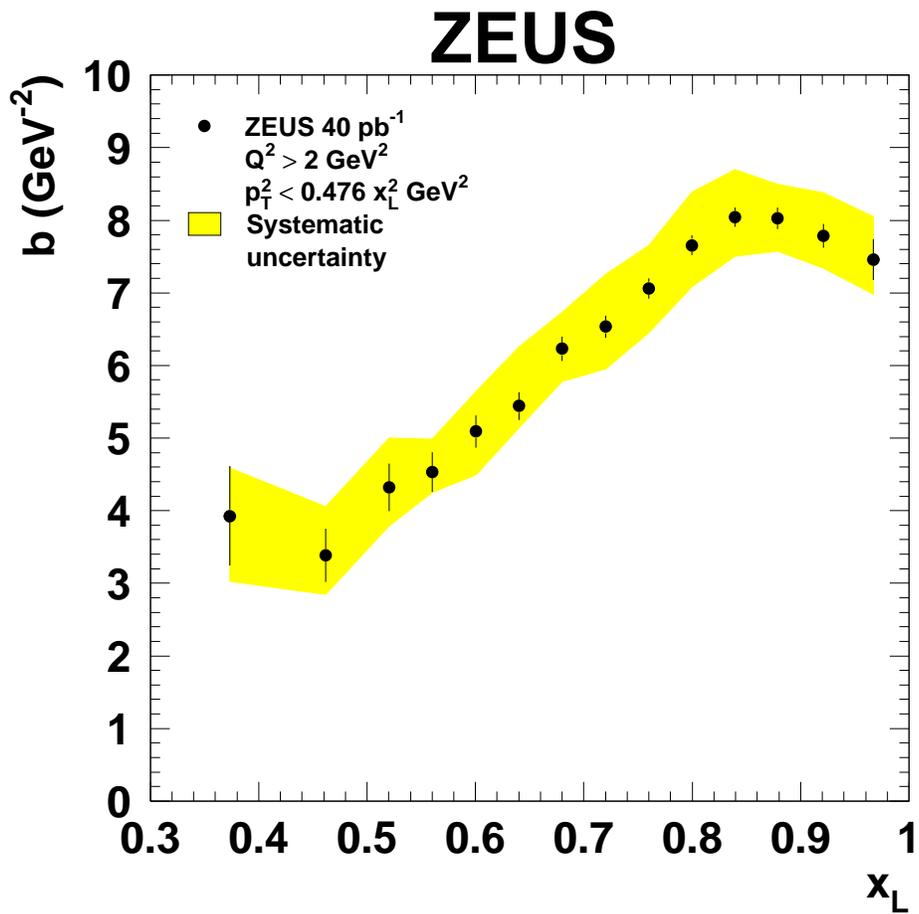,width=12cm}
\caption{
The exponential slopes $b$ versus $x_L$ from fits of the $p_T^2$ distributions
in DIS 
to the form $d\sigma_{\rm LN}/dp_T^2 = a \exp(-bp_T^2)$
over the $p_T^2$ ranges shown in \fig{ptsq_dis}.
The error bars show the statistical uncertainties;
the shaded band shows the systematic uncertainties.
}
\label{fig-bxlsys}
\end{figure}

\clearpage

\clearpage

\clearpage

\clearpage

\clearpage

\clearpage

\clearpage

\clearpage

\clearpage

\clearpage

\clearpage

\vfill\eject 
%
%

\begin{mcbibliography}{10}

\bibitem{pl:b384:388}
ZEUS \coll, M.~Derrick \etal,
\newblock Phys.\ Lett.{} {\bf B~384},~388~(1995)\relax
\relax
\bibitem{epj:c6:587}
H1 \coll, C.~Adloff \etal,
\newblock Eur.\ Phys.\ J.{} {\bf C~6},~587~(1999)\relax
\relax
\bibitem{np:b637:3}
ZEUS \coll, J.~Breitweg \etal,
\newblock Nucl.\ Phys.{} {\bf B~637},~3~(2002)\relax
\relax
\bibitem{np:b619:3}
H1 \coll, C.~Adloff \etal,
\newblock Nucl.\ Phys.{} {\bf B~619},~3~(2001)\relax
\relax
\bibitem{epj:c41:273}
H1 \coll, A.~Aktas \etal,
\newblock Eur.\ Phys.\ J.{} {\bf C~41},~273~(2005)\relax
\relax
\bibitem{np:b658:3}
ZEUS \coll, J.~Breitweg \etal,
\newblock Nucl.\ Phys.{} {\bf B~658},~3~(2003)\relax
\relax
\bibitem{np:b596:3}
ZEUS \coll, J.~Breitweg \etal,
\newblock Nucl.\ Phys.{} {\bf B~596},~3~(2001)\relax
\relax
\bibitem{pl:b590:143}
ZEUS \coll, S.~Chekanov \etal,
\newblock Phys.\ Lett.{} {\bf B~590},~143~(2004)\relax
\relax
\bibitem{pl:b610:199}
ZEUS \coll, S.~Chekanov \etal,
\newblock Phys.\ Lett.{} {\bf B~610},~199~(2005)\relax
\relax
\bibitem{BISH}
M.~Bishari,
\newblock Phys.\ Lett.{} {\bf B~38},~510~(1972)\relax
\relax
\bibitem{FMS}
L.L.~Frankfurt, L.~Mankiewicz and M.I.~Strikman,
\newblock Z.\ Phys.{} {\bf A~334},~343~(1989)\relax
\relax
\bibitem{GKS}
K.~Golec-Biernat, J.~Kwiecinski and A.~Szczurek,
\newblock Phys.\ Rev.{} {\bf D~56},~3955~(1997)\relax
\relax
\bibitem{pl:b338:363}
H.~Holtmann \etal,
\newblock Phys. Lett.{} {\bf B~338},~363~(1994)\relax
\relax
\bibitem{KPP}
B.~Kopeliovich, B.~Povh and I.~Potashnikova,
\newblock Z.\ Phys.{} {\bf C~73},~125~(1996)\relax
\relax
\bibitem{PSI}
M.~Przybycien, A.~Szczurek and G.~Ingelman,
\newblock Z.\ Phys.{} {\bf C~74},~509~(1997)\relax
\relax
\bibitem{MST}
W.~Melnitchouk, J.~Speth and A.W.~Thomas,
\newblock Phys.\ Rev.{} {\bf D~59},~014033~(1998)\relax
\relax
\bibitem{NSSS}
N.N. Nikolaev et al.,
\newblock Phys.\ Rev.{} {\bf D~60},~014004~(1999)\relax
\relax
\bibitem{SNS}
A.~Szczurek, N.N.~Nikolaev and J.~Speth,
\newblock Phys.\ Lett.{} {\bf B~428},~383~(1998)\relax
\relax
\bibitem{pr:188:2159}
J.~Benecke \etal,
\newblock Phys. Rev.{} {\bf 188},~2159~(1969)\relax
\relax
\bibitem{pr:d50:590}
T.T.~Chou and C.N.~Yang,
\newblock Phys. Rev.{} {\bf D~50},~590~(1994)\relax
\relax
\bibitem{hep-ph-9708290}
N.N.~Nikolaev, J.~Speth and B.G.~Zakharov,
\newblock Preprint \mbox{KFA-IKP(TH)-1997-17} (\mbox{hep-ph/9708290}), KFA-IKP,
  1997\relax
\relax
\bibitem{epj:a7:109}
U.~D'Alesio and H.J.~Pirner,
\newblock Eur. Phys. J.{} {\bf A~7},~109~(2000)\relax
\relax
\bibitem{epj:c47:385}
A.B.~Kaidalov \etal,
\newblock Eur. Phys. J.{} {\bf C~47},~385~(2006)\relax
\relax
\bibitem{Khoze:2006hwdo}
V.A.~Khoze, A.D.~Martin and M.G.~Ryskin,
\newblock Preprint \mbox{IPPP-06-36, DCPT-06-72} (\mbox{hep-ph/0606213}),
  2006\relax
\relax
\bibitem{zeus:1993:bluebook}
ZEUS \coll, U.~Holm~(ed.),
\newblock {\em The {ZEUS} Detector}.
\newblock Status Report (unpublished), DESY (1993),
\newblock available on
  \texttt{http://www-zeus.desy.de/bluebook/bluebook.html}\relax
\relax
\bibitem{nim:a309:77}
M.~Derrick \etal,
\newblock Nucl.\ Inst.\ Meth.{} {\bf A~309},~77~(1991)\relax
\relax
\bibitem{nim:a309:101}
A.~Andresen \etal,
\newblock Nucl.\ Inst.\ Meth.{} {\bf A~309},~101~(1991)\relax
\relax
\bibitem{nim:a321:356}
A.~Caldwell \etal,
\newblock Nucl.\ Inst.\ Meth.{} {\bf A~321},~356~(1992)\relax
\relax
\bibitem{nim:a336:23}
A.~Bernstein \etal,
\newblock Nucl.\ Inst.\ Meth.{} {\bf A~336},~23~(1993)\relax
\relax
\bibitem{desy-92-066}
J.~Andruszk\'ow \etal,
\newblock Preprint \mbox{DESY-92-066}, DESY, 1992\relax
\relax
\bibitem{zfp:c63:391}
ZEUS \coll, M.~Derrick \etal,
\newblock Z.\ Phys.{} {\bf C~63},~391~(1994)\relax
\relax
\bibitem{acpp:b32:2025}
J.~Andruszk\'ow \etal,
\newblock Acta Phys.\ Pol.{} {\bf B~32},~2025~(2001)\relax
\relax
\bibitem{nim:a354:479}
S.~Bhadra \etal,
\newblock Nucl.\ Inst.\ Meth.{} {\bf A~354},~479~(1995)\relax
\relax
\bibitem{nim:a394:121}
ZEUS \coll, FNC group, S.~Bhadra \etal,
\newblock Nucl.\ Inst.\ Meth.{} {\bf A~394},~121~(1997)\relax
\relax
\bibitem{proc:calor97:295}
ZEUS FNC Group, S.~Bhadra et al.,
\newblock {\em Proc.\ of the Seventh International Conference on calorimetry in
  High Energy Physics, Tuscon, Arizona, November 1997}, E.~Cheu et al.~(ed.),
  p.~295.
\newblock World Scientific, Singapore (1998)\relax
\relax
\bibitem{proc:hera:1991:23}
S.~Bentvelsen, J.~Engelen and P.~Kooijman,
\newblock {\em Proc.\ Workshop on Physics at {HERA}}, W.~Buchm\"uller and
  G.~Ingelman~(eds.), Vol.~1, p.~23.
\newblock Hamburg, Germany, DESY (1992)\relax
\relax
\bibitem{proc:epfacility:1979:391}
F.~Jacquet and A.~Blondel,
\newblock {\em Proceedings of the Study for an $ep$ Facility for {Europe}},
  U.~Amaldi~(ed.), p.~391.
\newblock Hamburg, Germany (1979).
\newblock Also in preprint \mbox{DESY 79/48}\relax
\relax
\bibitem{np:b627:3}
ZEUS \coll, S.~Chekanov \etal,
\newblock Nucl.\ Phys.{} {\bf B~627},~3~(2002)\relax
\relax
\bibitem{cpc:86:147}
H.~Jung,
\newblock Comp.\ Phys.\ Comm.{} {\bf 86},~147~(1995)\relax
\relax
\bibitem{pl:b366:371}
A.~Edin, G.~Ingelman and J.~Rathsman,
\newblock Phys.\ Lett.{} {\bf B~366},~371~(1996)\relax
\relax
\bibitem{zfp:c73:253}
ZEUS \coll, M.~Derrick \etal,
\newblock Z.\ Phys.{} {\bf C~73},~253~(1997)\relax
\relax
\bibitem{cpc:43:367}
T.~Sj\"ostrand and M.~Bengtsson,
\newblock Comp.\ Phys.\ Comm.{} {\bf 43},~367~(1987)\relax
\relax
\bibitem{cpc:101:108}
G.~Ingelman, A.~Edin and J.~Rathsman,
\newblock Comp.\ Phys.\ Comm.{} {\bf 101},~108~(1997)\relax
\relax
\bibitem{epj:c12:375}
CTEQ \coll, H.L.~Lai \etal,
\newblock Eur.\ Phys.\ J.{} {\bf C~12},~375~(2000)\relax
\relax
\bibitem{cpc:71:15}
L.~L\"onnblad,
\newblock Comp.\ Phys.\ Comm.{} {\bf 71},~15~(1992)\relax
\relax
\bibitem{cpc:143:100}
H.~Jung,
\newblock Comp. Phys. Comm.{} {\bf 143},~100~(2002)\relax
\relax
\bibitem{cpc:82:74}
T.~Sj\"ostrand,
\newblock Comp.\ Phys.\ Comm.{} {\bf 82},~74~(1994)\relax
\relax
\bibitem{phojetref}
R.~Engel, M.A.~Braun, C.~Pajares and J.~Ranft,
\newblock Z.\ Phys.{} {\bf C~74},~687~(1997)\relax
\relax
\bibitem{zfp:c76:613}
H1 \coll, C.~Adloff \etal,
\newblock Z.\ Phys.{} {\bf C~76},~613~(1997)\relax
\relax
\bibitem{zfp:c53:651}
M.~Gl\"uck, E.~Reya and A.~Vogt,
\newblock Z.\ Phys.{} {\bf C~53},~651~(1992)\relax
\relax
\bibitem{epj:c7:609}
ZEUS \coll, J.~Breitweg \etal,
\newblock Eur.\ Phys.\ J.{} {\bf C~7},~609~(1999)\relax
\relax
\bibitem{np:b713:3}
ZEUS \coll, S.~Chekanov \etal,
\newblock Nucl.\ Phys.{} {\bf B~713},~3~(2005)\relax
\relax
\bibitem{Khoze:2006reg}
V.A.~Khoze, A.D.~Martin and M.G.~Ryskin,
\newblock Preprint \mbox{IPPP-06-65, DCPT-06-130} (\mbox{hep-ph/0609312}),
  2006\relax
\relax
\bibitem{proc:hera:1991:1419}
G.A.~Schuler and H.~Spiesberger,
\newblock {\em Proc.\ Workshop on Physics at {HERA}}, W.~Buchm\"uller and
  G.~Ingelman~(eds.), Vol.~3, p.~1419.
\newblock Hamburg, Germany, DESY (1991)\relax
\relax
\bibitem{pr:d55:1280}
H.L.~Lai \etal,
\newblock Phys.\ Rev.{} {\bf D~55},~1280~(1997)\relax
\relax
\end{mcbibliography}
\end{document}